\documentstyle[epsfig]{mn}

\newif\ifAMStwofonts

\ifoldfss
  \ifCUPmtlplainloaded \else
    \NewTextAlphabet{textbfit} {cmbxti10} {}
    \NewTextAlphabet{textbfss} {cmssbx10} {}
    \NewMathAlphabet{mathbfit} {cmbxti10} {} 
    \NewMathAlphabet{mathbfss} {cmssbx10} {} 
  \fi
  \ifAMStwofonts
    \ifCUPmtlplainloaded \else
      \NewSymbolFont{upmath} {eurm10}
      \NewSymbolFont{AMSa} {msam10}
      \NewMathSymbol{\upi}     {0}{upmath}{19}
      \NewMathSymbol{\umu}     {0}{upmath}{16}
      \NewMathSymbol{\upartial}{0}{upmath}{40}
      \NewMathSymbol{\leqslant}{3}{AMSa}{36}
      \NewMathSymbol{\geqslant}{3}{AMSa}{3E}

      \let\leq=\leqslant \let\le=\leqslant
       \let\ge=\geqslant
    \fi
  \fi
\fi 

\ifnfssone
  \newmathalphabet{\mathit}
  \addtoversion{normal}{\mathit}{cmr}{m}{it}
  \addtoversion{bold}{\mathit}{cmr}{bx}{it}
  \newmathalphabet{\mathbfit} 
  \addtoversion{normal}{\mathbfit}{cmr}{bx}{it}
  \addtoversion{bold}{\mathbfit}{cmr}{bx}{it}
  \newmathalphabet{\mathbfss} 
  \addtoversion{normal}{\mathbfss}{cmss}{bx}{n}
  \addtoversion{bold}{\mathbfss}{cmss}{bx}{n}
  \ifAMStwofonts
    \ifCUPmtlplainloaded \else
      \UseAMStwoboldmath
      \makeatletter
      \new@mathgroup\upmath@group
      \define@mathgroup\mv@normal\upmath@group{eur}{m}{n}
      \define@mathgroup\mv@bold\upmath@group{eur}{b}{n}
      \edef\UPM{\hexnumber\upmath@group}
      \new@mathgroup\amsa@group
      \define@mathgroup\mv@normal\amsa@group{msa}{m}{n}
      \define@mathgroup\mv@bold\amsa@group{msa}{m}{n}
      \edef\AMSa{\hexnumber\amsa@group}
      \makeatother
      \mathchardef\upi="0\UPM19
      \mathchardef\umu="0\UPM16
      \mathchardef\upartial="0\UPM40
      \mathchardef\leqslant="3\AMSa36
      \mathchardef\geqslant="3\AMSa3E

      \let\leq=\leqslant \let\le=\leqslant
       \let\ge=\geqslant
    \fi
  \fi
\fi 

\ifnfsstwo
  \DeclareMathAlphabet{\mathbfit}{OT1}{cmr}{bx}{it}
  \SetMathAlphabet\mathbfit{bold}{OT1}{cmr}{bx}{it}
  \DeclareMathAlphabet{\mathbfss}{OT1}{cmss}{bx}{n}
  \SetMathAlphabet\mathbfss{bold}{OT1}{cmss}{bx}{n}
  \ifAMStwofonts
    \ifCUPmtlplainloaded \else
      \DeclareSymbolFont{UPM}{U}{eur}{m}{n}
      \SetSymbolFont{UPM}{bold}{U}{eur}{b}{n}
      \DeclareSymbolFont{AMSa}{U}{msa}{m}{n}
      \DeclareMathSymbol{\upi}{0}{UPM}{"19}
      \DeclareMathSymbol{\umu}{0}{UPM}{"16}
      \DeclareMathSymbol{\upartial}{0}{UPM}{"40}
      \DeclareMathSymbol{\leqslant}{3}{AMSa}{"36}
      \DeclareMathSymbol{\geqslant}{3}{AMSa}{"3E}

      \let\leq=\leqslant \let\le=\leqslant
       \let\ge=\geqslant
    \fi
  \fi
\fi 

\ifCUPmtlplainloaded \else
  \ifAMStwofonts \else 
    \def\upi{\pi}
    \def\umu{\mu}
    \def\upartial{\partial}
  \fi
\fi

\title{Effects of the integrated galactic IMF on the chemical evolution of the solar neighbourhood}
\author[F. Calura, S. Recchi, F. Matteucci, P. Kroupa]
       {F. Calura$^{1,2}$\thanks{E-mail: fcalura@oats.inaf.it}, S. Recchi$^{3}$, F. Matteucci$^{1,2}$, P. Kroupa$^{4}$\\
        (1) Dipartimento di Astronomia-Universit\'a di Trieste, Via G.B. Tiepolo
	11, 34131 Trieste, Italy\\
	(2) INAF, Osservatorio Astronomico di Trieste, via G.B. Tiepolo
	11, 34131 Trieste, Italy\\
        (3) Institute of Astronomy, Vienna University,
                T\"urkenschanzstrasse 17, A-1180, Vienna, Austria\\
        (4) Argelander Institute for Astronomy, Bonn University,
              Auf dem H\"ugel 71, 53121 Bonn, Germany
	 }
	
\date{Accepted ---- .
      Received ---- ;
      in original form ----}

\pagerange{\pageref{firstpage}--\pageref{lastpage}}
\pubyear{----}

\begin{document}

\maketitle

\label{firstpage}
\begin{abstract}
The  initial  mass  function  determines  the  fraction  of  stars  of
different intial mass born per  stellar generation.  In this paper, we
test  the effects  of the  integrated galactic  initial  mass function
(IGIMF)  on the chemical  evolution of  the solar  neighbourhood.  The
IGIMF (Weidner \& Kroupa 2005) is computed from the combination of the
stellar intial mass  function (IMF), i.e. the mass  function of single
star clusters,  and the embedded  cluster mass function, i.e.  a power
law with index $\beta$.  By taking into account also the fact that the
maximum achievable stellar mass is a function of the total mass of the
cluster, the  IGIMF becomes  a time-varying  IMF which
depends on  the star formation rate.   We applied this  formalism to a
chemical evolution model for  the solar neighbourhood and compared the
results obtained  by assuming three  possible values for  $\beta$ with
the  results obtained by  means of  a standard,  well-tested, constant
IMF.  In general, a lower  absolute value of $\beta$ implies a flatter
IGIMF,  hence  a larger  number  of  massive  stars and  larger  metal
ejection rates.  This translates into  higher type Ia and II supernova
rates,  higher mass  ejection rates  from massive  stars and  a larger
amount  of  gas  available  for  star formation,  coupled  with  lower
present-day   stellar  mass  densities.    Lower  values   of  $\beta$
correspond  also  to  higher  metallicities and  higher  [$\alpha$/Fe]
values at  a given  metallicity. We consider  a large set  of chemical
evolution  observables and test  which value  of $\beta$  provides the
best  match  to  all  of  these  constraints.   We  also  discuss  the
importance  of  the  present  day  stellar  mass  function  (PDMF)  in
providing a way to  disentangle among various assumptions for $\beta$.
Our results indicate  that the model adopting the  IGIMF computed with
$\beta  \simeq 2 $  should be  considered  the best  since it  allows us  to
reproduce the  observed PDMF and to  account for most  of the chemical
evolution constraints considered in this work.
\end{abstract} 


\section{Introduction} 
The initial stellar mass function is one of the major ingredients of chemical 
evolution models. Moreover, observed 
chemical abundances allow one to put robust 
constraints on both the normalization and the slope of the initial mass function (IMF; Chiappini et al. 2000; Romano et al. 2005). 
The environment 
providing most observational constraints for chemical evolution studies is the solar neighbourhood (S. N. hereinafter), 
for which a large set of observables are available. These observables include diagrams of abundance 
ratios versus metallicity, particularly useful when they involve two elements synthesised by stars on different timescales. 
An example is the 
[$\alpha$/Fe] vs [Fe/H] diagram, since $\alpha$ elements are produced by massive stars on short ($<0.03$ Gyr) timescales, 
while type Ia supernovae (SNe) produce mostly Fe on timescales spanning from 0.03 Gyr up to one Hubble time (Matteucci 2001). 
This diagnostic is a strong function of the IMF, but 
depends also on the assumed star formation history (Matteucci 2001; Calura et al. 2009). 
Another fundamental constraint is the metallicity distribution of living stars, 
which provides us with fundamental information on the IMF and on the infall history of the studied system. 
Another  diagnostic, depending both on the IMF and the past 
star formation history, is the present-day mass function, which represents the mass function of living stars observed now in the Solar 
Vicinity (Elmegreen \& Scalo 2006). 
Other important observables useful for chemical evolution studies include the type Ia and type II SN rates, as well as the surface density 
of stars and gas, depending on the IMF and on the rate at which the gas has been processed into stars and remnants in the past, i.e. on the SFR. 

In a previous paper, Recchi et al. (2009) 
considered a star-formation dependent IMF, called the integrated galactic initial mass function (IGIMF), 
which originates from the combination of the stellar IMF within each star cluster and 
the embedded cluster mass function.  
\footnote{Embedded clusters are stellar clusters that are partially or fully 
encased in interstellar gas and dust within molecular clouds,
therefore often visible only in the infrared.  It is supposed that all
(or the large majority of) the stars form originally in embedded
clusters (Lada \& Lada 1991), but then they can loose their cocoon of
gas because of the feedback of O stars (see e.g. Boily \& Kroupa
2003a,b).  We will name therefore hereafter ``embedded clusters'' also
the clusters which have lost their envelope, but still retain all of their stars, 
to be consistent with the
terminology used in the original papers describing the IGIMF theory
(e.g. Kroupa \& Weidner 2003; Weidner \& Kroupa 2005).}  Within each
star cluster, the IMF can be well approximated
by a two-part power-law form, $\xi(m) \propto m^{-\alpha}$
(e.g. Pflamm-Altenburg, Weidner \& Kroupa 2007). Massey \& Hunter
(1998) have shown that for stellar masses $m>$ a few M$_\odot$, a
slope similar to the Salpeter (1955) index (i.e. $\alpha=2.35$) can
approximate well the IMF in clusters and OB associations for a wide
range of metallicities. Other studies have shown that the IMF flattens
out below $m$ $\sim$ 0.5 M$_\odot$ (Kroupa, Tout \& Gilmore 1993;
Chabrier 2003). On the other hand, the embedded cluster mass function
is well approximated by a single slope power law. This implies that
small embedded clusters are more numerous in galaxies and they lock up
most of of the stellar mass. However, the most massive stars tend to
form preferentially in massive clusters (Weidner \& Kroupa 2006).
The integrated IMF in galaxies, the IGIMF, 
is a function of the galactic star formation rate (SFR) and, as a consequence of the 
embedded cluster mass function, it 
is steeper than the stellar
IMF within each single star cluster (Kroupa \& Weidner 2003;
Weidner \& Kroupa 2005). 

Recchi et al. (2009) 
studied the effects of the IGIMF 
on the evolution of the SN rates in galaxies and
on the chemical evolution of elliptical galaxies, 
showing how the IGIMF naturally accounts for the relation between the [$\alpha$/Fe] and the stellar velocity dispersion 
observed in local elliptical galaxies. 
In this paper, we consider the effects of the IGIMF on the chemical evolution of the solar neighbourhood. 
As already stressed, the advantage of this approach is the availability of a large set of observational constraints, 
useful to test the IGIMF and, most importantly, to constrain its main parameter, i.e.  the index $\beta$ 
of the power law expressing the embedded clusters mass function.
 
We will compare the results computed by means of a standard IMF, similar to the one by Scalo (1986), 
successful in reproducing most of the chemical evolution properties of the Solar Neighbourhood, 
with the results computed by means of the IGIMF. 
This Paper is organized as follows. In Section 2 
we describe the IMF used in standard chemical evolution models and the formalism behind the IGIMF. 
In Section 3
we present a brief description of the chemical evolution model of the Solar Neighbourhood.
In Sect. 4 we present our results and in Sect. 5 we draw our conclusions. 

\section{The Initial Mass Function}
\subsection{The standard initial mass function}
\label{standard}
$\xi_{std}(m)$ is 
the initial mass function assumed in the standard chemical 
evolution model used in this paper and is a two-slope power law, defined in number as:
\begin{equation}
     \xi_{std}(m) = \left\{ \begin{array}{l l}
                                      0.19\, \cdot m^{-2.35} & 
			     \qquad {\mathrm{if}} \; m < 2 \, M_\odot \\
			              0.24\, \cdot m^{-2.70} &
			     \qquad {\mathrm{if}} \; m > 2 \, M_\odot, \\
                                      \end{array} \right.
\label{standard_imf} 
\end{equation}
In the remainder of the paper, we will refer to the IMF of Eq.~\ref{standard_imf} as to the \emph{Standard} IMF.  
This equation represents  a simplified two-slope approximation of  the actual  
Scalo (1986) IMF, similarly to what is done in Matteucci \& Fran\c cois 
(1989). 
Our basic IMF is assumed to be constant in space and time and normalized in mass to unity in the mass interval $0.1 -100 M_{\odot}$, i.e.:
\begin{equation}
\int m \xi_{std}(m) dm = 1.
\label{norm}
\end{equation}
Various papers have shown that, 
by assuming this IMF it is possible to reproduce a large number of observational 
constraints for the solar neighbourhood (Chiappini et al. 2001; Romano et al. 2005). 
By means of our chemical evolution model, we will present predictions for various 
observables computed by assuming the standard IMF. These predictions will be tested 
against observed quantities and will be compared to results computed assuming the integrated 
galactic IMF, which is the subject of the following section. 

\subsection{The integrated galactic initial mass function}
\label{sec_igimf} 
The IGIMF theory has been described in detail in previous papers (Kroupa \&
Weidner 2003; Weidner \& Kroupa 2005; Recchi, Calura \& Kroupa 2009).  Here we
briefly summarize its main assumptions and features.  The IGIMF theory is
based on the assumption that all the stars in a galaxy form in star clusters.
Within each embedded cluster, the stellar IMF has the canonical form $\xi(m) =
k m^{-\alpha}$, with $\alpha = 1.3$ for $\sim$ 0.1 M$_\odot \le$ $m$ $<$ 0.5
M$_\odot$ and $\alpha = 2.35$ (i.e. the Salpeter slope) for 0.5 M$_\odot \le$
$m$ $< m_{\rm max}$.  The upper mass $m_{\rm max}$ depends on the mass of the
embedded cluster $M_{\rm ecl}$ simply because small clusters do not have
enough mass to produce very massive stars. 

On the other hand, star clusters are also apparently distributed according to
a single-slope power law, $\xi_{\rm ecl} \propto M_{\rm ecl}^{-\beta}$ (Zhang
\& Fall 1999; Lada \& Lada 2003).  In this work we have assumed 3 possible
values of $\beta$: 1.00, 2.00 and 2.35 .  
By convolving the stellar IMF with the distribution of embedded
clusters we obtain the IGIMF, namely the IMF integrated over the whole
population of embedded clusters forming in a galaxy as a function of the star formation rate 
$\psi(t)$:

\begin{equation}
\xi_{\rm IGIMF}(m;{\psi} (t)) = 
\int_{M_{\rm ecl, min}}^{M_{\rm ecl, max} ({\psi} (t))} 
\hspace{-0.6cm}\xi (m \leq m_{\rm max}) \xi_{\rm ecl} (M_{\rm ecl}) 
dM_{\rm ecl}, 
\end{equation}
where $M_{\rm ecl, min}$ and $M_{\rm ecl, max} (\psi (t))$ are the
minimum and maximum possible masses of the embedded clusters in a population
of clusters, respectively, and $m_{\rm max} = m_{\rm max} (M_{\rm ecl})$.  For $M_{\rm ecl,
  min}$ we take 5 M$_\odot$ (the mass of a Taurus-Auriga aggregate, which is
arguably the smallest star-forming "cluster" known).  The upper mass of the
embedded cluster population depends instead on the SFR and that makes the whole IGIMF dependent on $\psi$.  The correlation
between $M_{\rm ecl, max}$ and SFR has been determined observationally (Larsen
\& Richtler 2000; Weidner et al. 2004) and results from
the sampling of clusters from the embedded cluster mass function given the
amount of gas mass being turned into stars per unit time (Weidner et
al. 2004).

\begin{figure*}
\centering
\vspace{0.001cm}
\epsfig{file=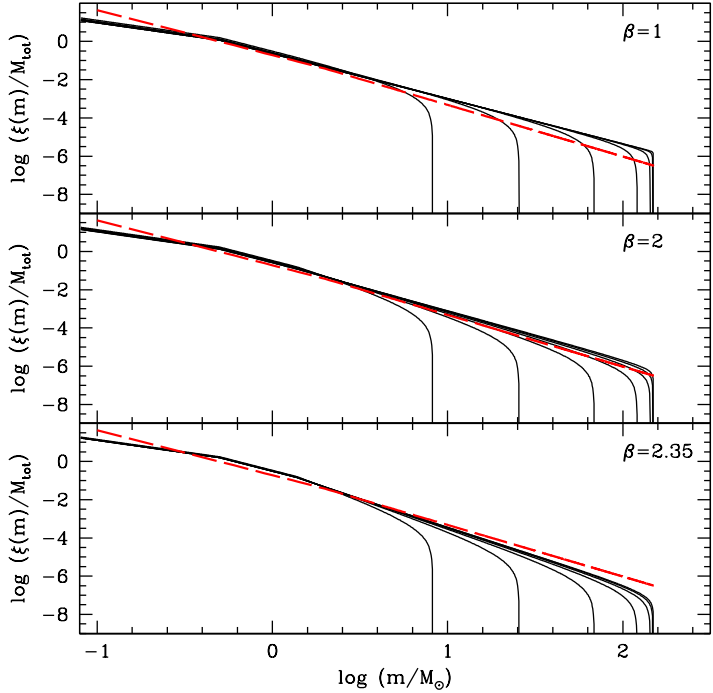,height=10cm,width=10cm}
\caption{IGIMFs for different cluster mass functions distributions. 
Upper panel: $\beta=1$; central
panel: $\beta=2$; lower panel: $\beta=2.35$. 
In 
each panel we have considered 7 possible values of SFRs, 
ranging
from $10^{-−4}$ $M_{\odot} \, yr^{-1}$ (lowermost solid lines) to 100 $M_{\odot} \, yr^{-1}$ (uppermost
solid lines), equally spaced in logarithm. The thick dashed line is the standard IMF used in this work (see Sect.~\ref{standard}). 
The IGIMFs and the standard IMF are all normalized through the equation $\int m \xi (m) dm=M_{tot}=1M_{\odot}$.
}
\label{igimf}
\end{figure*}

In Fig.~\ref{igimf}, we show the IGIMF as a function of the SFR for the three values of $\beta$ considered 
in this work, compared to our standard IMF. 
The IGIMFs are characterized by a nearly uniform decline, which follows
approximately a power law, and a sharp cutoff when $m$ gets close to $m_{\rm
  max}$.  Of course, the steepest distribution of embedded cluster (in our
case the model with $\beta=2.35$) produces also the steepest IGIMF because this distribution  is
biased towards embedded clusters of low mass, therefore the probability of
finding high mass stars in this cluster population is lower.  Moreover, the
dependence of the IGIMF on the SFR is strong for SFR $\le$ 1 M$_\odot$
yr$^{-1}$ whereas it is very mild for SFR $\ge$ 1 M$_\odot$ yr$^{-1}$ (see
Recchi et al. 2009).  This is due to the fact that for SFRs larger than $\sim$
1 M$_\odot$ yr$^{-1}$ the maximum possible mass of the embedded cluster is
very high, therefore it is always possible to sample massive stars in the
whole galaxy up to a mass very close to the empirical limit (which is
assumed to be 150 M$_\odot$; see Weidner \& Kroupa 2005).  All the IGIMFs are normalized in mass  to unity 
as the standard IMF (see eq.~\ref{norm}).\\

The standard IMF is steeper in the low mass range, i.e. 
for stellar masses $m \le$ 0.5 M$_\odot$. 
These masses do not contribute to the
chemical enrichment of the Galaxy, since they have lifetimes larger than
the Hubble time. However, their distribution remains
unchanged during the evolution of the Galaxy and it can be constrained
by analysing the present-day mass function (see Sect.~\ref{PDMF}).


\section{The Chemical Evolution model} 
\label{chem}
The adopted chemical evolution model is calibrated in order to reproduce a large set of observational 
constraints for the Milky Way galaxy (Chiappini et al. 2001). 
The Galactic disc is approximated by several independent rings, 
2 kpc wide, without exchange of matter between them. The Milky Way 
is  assumed to form as a result of two main infall episodes.  
During the first episode, the halo and the thick disc are formed.
During the second episode, a slower infall
of external gas forms the thin disc with the gas accumulating faster in the inner than in the outer
region ("inside-out" scenario, Matteucci \& Fran\c cois 1989). The process of disc formation is much longer than the
halo 
and bulge formation, with time scales varying from $\sim2$ Gyr in the inner disc to $\sim7$ Gyr in the solar region
and up to $20$ Gyr in the outermost disc. 
In this paper, we are interested in the effects of a time-variyng IMF in the Solar Neighbourhood. 
For this purpose, we focus on a ring located at 8 kpc from the Galactic centre, 2 kpc wide. 

 For a single-phase gas, 
the chemical evolution of a given chemical element \emph{i} is computed through the following equation:
\begin{eqnarray}
{d G_i (t) \over d t} = -\dot{\sigma}_{*} (t) X_i (t) \nonumber \\
 +\int_{M_L}^{M_{B_m}} \dot{\sigma}_{*} (t-\tau_m) Q_{\rm mi}(t-\tau_m) \xi^{'} (m,t) dm\,  \nonumber \\
 + A_{Ia}\int_{M_{B_m}}^{M_{B_M}} \int_{\rm\mu_{\rm min}}^{0.5}
 f(\mu) Q_{\rm mi}(t-\tau_{\rm m_2})\xi^{'} (m,t)  \dot{\sigma}_{*} (t-\tau_{\rm m_2}) d\mu dm \,  \nonumber \\
  +(1-A_{Ia})\, \int_{\rm M_{\rm B_m}}^{M_{\rm B_M}} \dot{\sigma}_{*} (t-\tau_m)\,  
 Q_{\rm mi}(t-\tau_m) \xi^{'} (m,t) dm  \,  \nonumber \\
  +\int_{\rm M_{\rm B_M}}^{M_{U}(t)} \dot{\sigma}_{*} (t-\tau_m) Q_{\rm mi}(t-\tau_m) \xi^{'} (m,t) dm\,  + (\frac{d G_i (t)}{d t})_{\rm inf}, \nonumber \\
\label{eq_chem}
\end{eqnarray}
where $G_{i}(t)=\sigma_{g}(t)X_{i}(t)/\sigma_{tot}$ is the gas surface mass density in 
the form of an element $i$ normalized to a total surface mass density  
$\sigma_{tot}$, and $G(t)= \sigma_{g}(t)/\sigma_{tot}$ is the total fractional 
mass of gas present in the galaxy at the time $t$.
$X_i (t)$ is the abundance by mass (or mass fraction) of the element \emph{i}. 
The quantity $\dot{\sigma}_{*} (t)$ is the surface SFR density. In general, this quantity is 
a function of the galactic radius $r$:
\begin{equation}
\dot{\sigma}_{*}(r,t) = \nu [\frac{\sigma(r,t)}{\sigma(r_{\odot},t)}]^{2(k-1)} [\frac{\sigma(r,t_{Gal})}{\sigma(r,t)}]^{k-1}
\sigma^{k}_{g}(r,t)
\end{equation} 
(Chiappini et al. 1997), where $\nu$ is the SF efficiency, $\sigma(r,t)$ is the total mass 
(gas + stars) surface density at a radius $r$ and time $t$ ($t_{Gal}=14$ Gyr is the present age of the Milky Way), 
$\sigma(r_{\odot},t)$ is the total surface mass density in the solar 
region and $\sigma_{g}(r,t)$ is the ISM surface mass 
density. In this paper, we focus on the solar neighbouhod and we assume $r=r_{\odot}=8 kpc$. 
For the gas density exponent $k$ a value of 1.5 has been assumed 
by Chiappini et al. (1997) in order to ensure a good
fit to the observational constraints for a large set of local spirals (Kennicutt 1998). 
The efficiency of SF is set to $\nu=1$ Gyr$^{-1}$, and
becomes zero when the gas surface density drops below a certain 
critical threshold. 
For the SF, we adopt a threshold gas 
density $\sigma_{th}\sim 7 M_{\odot} pc^{-2}$ in the disc as suggested by 
Kennicutt (1989). 

The second term in Eq.~\ref{eq_chem} is the rate at which each element is restored into the ISM 
by single stars with masses in the range $M_{L}$ - $M_{B_m}$, where $M_{L}$ is the minimum mass contributing, 
at a given time $t$, to chemical enrichment 
(the minimum is $0.8 M_{\odot}$) 
and $M_{B_m}$ is the minimum mass allowed for binary systems giving rise to type Ia SN ($3 M_{\odot}$, Matteucci \& Greggio 1986). 
The quantity  
\begin{equation}
\xi^{'}(m, t) = m \, \xi(m, t)
\end{equation}
 is the initial mass function in mass and in the standard case is given by $\xi^{'}_{std}= m \, \xi_{std}$, 
described in Sect.  \ref{standard} and is constant in time, otherwise it is a function of the 
initial stellar mass $m$ and of the time $t$ and is computed with the method described 
in Sect. \ref{sec_igimf}. 
The quantities $Q_{mi}(t-\tau_m)$ (where $\tau_m$ is the lifetime of a star of mass $m$) contain all
the information about stellar nucleosynthesis for elements either produced or destroyed inside 
stars or both (Talbot and Arnett 1971; Matteucci 2001).  
The third term represents the enrichment due to binaries which become type Ia SNe, i.e. all the binary systems with total mass 
between $M_{B_m}$ and $M_{B_M}=16 M_{\odot}$. For the type Ia SN progenitor model, the Single Degenerate (SD) scenario is assumed, 
where a C-O white dwarf explodes by C-deflagration mechanism 
after having reached the Chandrasekhar mass ($1.44 M_{\odot}$), owing to progressive 
mass accretion from a non-degenerate companion (Whelan \& Iben 1973). 
This model is still one of the best to reproduce the majority of the 
properties of local galaxies (Matteucci et al. 2006; Calura \& Matteucci 2006; Matteucci et al. 2009). 
The parameter $A_{Ia}$ represents the unknown 
fraction of binary systems with the specific characteristics to become type Ia SNe in the range 3-16 $M_{\odot}$
and is fixed by reproducing the observed 
present time SN Ia rate (Calura \& Matteucci 2006). 
In this third term, both quantities $\dot{\sigma}_{*}$ and $Q_{mi}$ refer to the time $t-t_{m_2}$, where $t_{m_2}$ indicates the lifetime of the 
secondary star of the binary system, which regulates the explosion timescale. 
$\mu=M_{2}/M_{B}$ is the ratio between the mass of the secondary component $M_{2}$ 
and the total mass of the binary system $M_{B}$, whereas 
$f(\mu)$ is the 
distribution function of this ratio. Statistical studies indicate that mass 
ratios close to $0.5$ are preferred, so the formula:\\
\begin{equation}
f(\mu)=2^{1+\gamma}(1+\gamma)\mu^{\gamma}
\end{equation}
is commonly adopted, with $\gamma=2$ (Matteucci \& Recchi 2001). 
$\mu_{min}$ is the minimum mass fraction contributing to the SNIa rate at the time $t$, and is given by\\ 
\begin{equation}
\mu_{min}=max \left \{ \frac{M_{2}(t)}{M_{B}}, \frac{M_{2}-0.5M_{B}}{M_{B}} \right \}.
\end{equation}
The fourth term represents the enrichment due to stars in the mass range $M_{B_m}$ - $M_{B_M}$ which are either single, or, 
if in binaries, do not produce a SN Ia event. In this mass range, all the stars with masses $m > 8 M_{\odot}$ will explode as 
type II SNe, which in our picture are assumed to originate from core collapse of single massive stars. 
The fifth term of Eq.~\ref{eq_chem} represents the enrichment of stars more massive than $M_{B_M}$, 
all of which explode as core collapse (i.e mostly type II, see Calura \& matteucci 2006) SNe. 
The upper mass limit contributing to chemical enrichment is $M_{U(t)}$ and is a function of time through 
the IMF, $\xi^{'} (m,t)$. We assume that the maximum value for $M_{U(t)}$ is 100 $M_{\odot}$. The assumption 
of a maximum stellar mass 150 $M_{\odot}$ would have a negligible effect on the results presented in this paper. 
Finally, the last term accounts 
for the infall  of external gas.

It is worth stressing that equation~\ref{eq_chem} differs substantially from the classic formalism used in most of the previous 
chemical 
evolution models (e.g. Matteucci \& Greggio 1986; Chiappini et al. 2001), which use an IMF constant in time, whereas in this case 
the IMF is a function of the SFR, which is a function of  cosmic time. Later on, we will see how the time variations  
of the IGIMF depend on the star formation history for various values of the parameter $\beta$.

The rate at which the thin disc is formed out of external matter is 
\begin{equation}
({d G_i (t) \over d t})_{\rm inf} = B(R)\,e^{- (t - t_{max})/\tau_{D}},
\end{equation} 
 where  $t_{max}$ is the time of maximum 
gas accretion onto the disc, corresponding to the end of the halo-thick disc 
phase and  it is equal to 1 Gyr. The quantity  $\tau_{D}$ is the timescale for mass 
accretion onto the thin disc component. Following Romano et al. (2000) and Chiappini et al. (2001), 
we assume that $\tau_{D}$ increases with increasing Galactic radius
\begin{equation}
\tau_{D}(R) = 1.033 \times R - 1.267.
\end{equation}
The constant $B(R_{\odot})$ is fixed in order to  reprodue the present-day total surface mass density (stars + gas) in the solar neighbourhood. 
Concerning the nucleosynthesis prescriptions, we assume the yields of Van den Hoeck \& Groenewegen (1997) for low and intermediate mass 
stars, the yields of Iwamoto et al. (1999) for type Ia SNe and those of Fran\c cois et al. (2004) for massive stars. 
In Table 1, we show the standard assumptions for the main parameters of our chemical evolution model.

\subsection{Observational data set}
In Table ~\ref{data}, we show the solar neighbourhood observables used in this paper, with their 
values and references.\\
Concerning the observed type Ia and type II SN rates, as observational data  
we use the  values of Capellaro (1996), expressed 
in $century^{-1}$ and valid for the whole MW disc. To compare them with the predictions,  
we divide the observed values by the area of the MW disc, which, assuming a radius of 
$\sim 15$ kpc, is roughly $S_{disc}\sim 10^9 pc^{2}$. 
In this way, for the type Ia and type II SN rates we obtain  
values of $0.003$ $pc^{-2}$Gyr$^{-1}$ and $0.012$ $pc^{-2} $Gyr$^{-1}$, respectively.

Concerning the stellar surface density of visible stars and remnants, 
as observational value we use an indirect estimate. 
In a recent paper, Weber \& de Boer (2009) discuss the uncertainty of the 
total local visible surface density. 
They provide a value of  $48 \pm 9$ $M_{\odot}/pc^{2}$. 
This value is in agreement with another recent estimate of Holmberg \& Flynn (2004), which found 
53 $M_{\odot}/pc^{2}$ in visible matter, and is 
consistent  with the value we achieve with our standard model, 50 $M_{\odot}/pc^{2}$.  

The local gas surface density is between 7 and 14  $M_{\odot}/pc^{2}$
(Kulkarni \& Heyles 1987, Dame 1993, Olling \& Merrifield 2001).
A reasonable value based on this data is 10.5 $\pm$ 3.5 $M_{\odot}/pc^{2}$. 
This value is also consistent with our estimate from the standard model. 
By combining these two
quantities, i.e. by subtracting the gas density from the total density and 
by combining the errors, 
for the local mass density in stars and remnants we obtain 37.5 $\pm$ 10$M_{\odot}/pc^{2}$. 
This estimate is compatible with previous values  based on the combination of the observed stellar 
mass density and the one in stellar remnants
(Gilmore et al. 1989, Mera et al. 1998).

\begin{table*}
\vspace{0cm}
\begin{flushleft}
\caption[]{Parameters for the standard model of the solar neighbourhood. }
\begin{tabular}{l|l}
\noalign{\smallskip}
\hline
\hline
\noalign{\smallskip}
 Parameter             & Adopted value \\ 
\noalign{\smallskip}
\hline
\noalign{\smallskip}
SF efficiency &   \\
$\nu$ $($Gyr$^ {-1})$     &  1  \\
\hline
Infall timescale for the      &  \\
thin disc $\tau_D$ $($Gyr$)$    & 7        \\
\hline
Gas density threshold &  \\
for SF                &   \\
$(M_{\odot}/pc^{2})$  & 7   \\
\hline
Fraction $A_{Ia}$ of binary systems  &  \\
originating type Ia SNe     & 0.04         \\
\hline
\hline
\end{tabular}
\end{flushleft}
\end{table*}
\begin{table*}
\vspace{0cm}
\begin{flushleft}
\caption[]{Solar neighbourhood observables used in this paper and references.}
\begin{tabular}{l|l|l}
\noalign{\smallskip}
\hline
\hline
\noalign{\smallskip}
 Observable             & value & Reference \\ 
\hline
\hline
SFR Surface density & $3.5 \pm 1.5$ $M_{\odot} \,  pc^{-2} \, Gyr^{-1}$  &  Rana (1991)\\
type Ia SNR         & $0.003 \pm 0.002$ $pc^{-2} \, Gyr^{-1}                   $ &  Cappellaro  (1996)\\
type II SNR         & $0.012 \pm 0.008$ $pc^{-2} \, Gyr^{-1}                $ &  Cappellaro  (1996)\\
Gas surface density &  $10.5 \pm 3.5$ $M_{\odot} \,  pc^{-2}$                     &    Kulkarni \& Heiles (1987)   \\
                    &                                                           &    Dame (1993)                 \\
                    &                                                           &  Olling \& Merrifield (2001)  \\
Stellar surface density          & $37.5  \pm 10 $ $M_{\odot} \,  pc^{-2}$         &   Weber \& de Boer (2009)  \\
(visible stars and remnants)     &                                               &                           \\
Stellar abundance ratios & - & various authors    \\
Stellar Metallicity distribution  & - & various authors    \\ 
Present-day mass function  & - & various authors    \\ 
\hline
\hline
\end{tabular}
\label{data}
\end{flushleft}
\end{table*}


\subsection{IGIMF and Star Formation Rate}

In Section 2, we have seen that the IGIMF is a function of the total 
SFR $\psi(t)$, expressed in $M_{\odot}/yr$. 
In particular, the upper mass limit  of the IGIMF is sensitive to the value of 
the star formation rate.  \\
Our chemical evolution code is designed to produce all physical 
quantities as surface mass densities, so the direct output of our code is 
a SFR surface density (the quantity $\dot{\sigma}_{*}$ defined in Sect.~\ref{chem}). 
To convert the SFR density into the appropriate units, we 
perform the following assumptions. 
The solar neighbourhood is described by a 2 kpc-wide ring located 
at the galactocentric distance of 8 kpc from the centre. 
The area of this ring is $S_{\odot}\sim 10^{8}$ pc$^2$. 
The SFR $\psi(t)$ can be calculated from the SFR surface density $\dot{\sigma}_{*}$ as 
\begin{equation}
\psi(t)= \dot{\sigma}_{*} \cdot S_{\odot} \cdot 10^{-9}.
\end{equation}       
In this way, we obtain for our standard model a present value 
of $0.26 M_{\odot}/yr$. 
If we consider that the local observed SFR density is $3.5 M_{\odot} pc^{-2} yr^{-1}$ and we convert 
this value into the same units, we obtain $0.35 M_{\odot}/yr$, consistent with the 
above estimate.

\section{Results} 
\label{results}
Our aim is to test the effects of the IGIMF on the chemical evolution of the S.N., taking into account 
a large set of available observational constraints. 
First, we will consider the effects of the SFR-dependent IGIMF on the predicted physical properties of the solar 
neighbourhood. 
Then, we will discuss all of the observables 
and the main parameters which can be tested by our analysis. 
Finally, we will see how a fine-tuning 
of the parameters considered in this work may allow us to derive constraints on the local IMF.

\subsection{The fitness test} 
\label{fitness}
In order to quantitatively compare our results with the observables
considered in this paper, we adopt the formula

\begin{equation}
fitness = {1 \over {1 + \delta}}~~;~~\delta = \Sigma_i {w(i)
[{\rm obs} (i) - {\rm theo} (i)]^2 \over 
\max\bigl\lbrace[{\rm obs} (i)]^2, [{\rm theo} (i)]^2\bigr\rbrace}
\end{equation}
\noindent
where, for the $i$-th value of each considered parameter, obs ($i$) and theo ($i$)
are the observed values and the predictions of the model, respectively
(see also R{\u u}{\v z}i{\v c}ka et al. 2007; Theis \& Kohle 2001).  We have also
introduced a weight $w (i)$ for each observable $i$ in order to give
each set of observables the same statistical weight.  To be more
precise: we have grouped the observables in 5 groups: (1) physical
quantities (SFR, SNR$_{II}$, SNR$_{Ia}$, $\Sigma_*$, $\Sigma_{gas}$);
(2) solar abundances; (3) average [$\alpha$/Fe] ([O/Fe], [Si/Fe])
ratios for each bin of [Fe/H]; (4) dN/d[Fe/H] (SMD) in each bin of [Fe/H];
(5) present-day mass function in each bin of mass.  We have chosen a weight such that $w (i)$
multiplied by the members of each group gives always the same number. 
This means that in the fitness calculation, the same weight is given to  
all the 5 groups of observables. 
Following this approach, each single observable in the physical
quantities group has the largest weight, and this occurs because the physical quanties 
group is the one having the smallest number of members (5). 
On the other hand,  each 
single member of the [$\alpha$/Fe] group, counting the highest member number (28),  has the
lowest one, but, as already stressed, the [$\alpha$/Fe] group as a whole has the same weight of the 
physical quantities group. 
Of course, the closer $fitness$ is to 1, the better the
model is in reproducing the observations.
In Fig.~\ref{fit}, we show the ``fitness'' quantity as 
a function of $\beta$ for all the models considered in this paper. 
The results obtained for each value of $\beta$ will be discussed separately. 

\subsection{The effects of the IGIMF on the properties of the solar neighbourhood}
\begin{figure*}
\centering
\vspace{0.001cm}
\epsfig{file=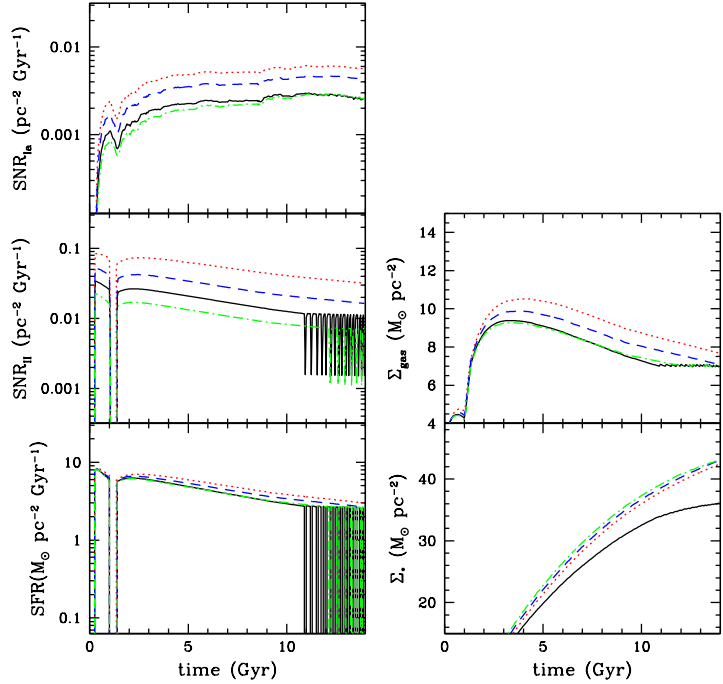,height=10cm,width=10cm}
\caption{From top-left, anti-clockwise: time evolution of the type Ia SNR, type II SN rate, SFR, surface density in stars and in gas computed with the 
standard IMF (solid lines) and with the IGIMF assuming $\beta=1$ (dotted lines), $\beta=2$ (dashed lines) and $\beta=2.35$ (dot-dashed lines). 
For all the models we have assumed the same value for the star formation efficiency, $\nu = 1 \, $Gyr$^{-1}$.}
\label{snr}
\end{figure*}

In Fig.~\ref{snr}  we show the effects of the standard IMF and of the IGIMF assuming three different values for 
$\beta$ on the calculated evolution of the star formation rate, of the type Ia and II SN rates and of 
the stellar and gas mass surface density.

All the models shown in Fig.~\ref{snr} are characherised by the same 
value for the star formation efficiency, $\nu=1$ Gyr$^{-1}$. In this way, it is possible to appreciate the effects of varying the IMF, keeping all 
the other parameters constant.

The most striking features of the standard model are the SF hiatus at 1 Gyr 
(see Chiappini et al. 1997, 2001) and the threshold-dominated SF after 10 Gyr.

The hiatus in the SF is due to the transition between 
the end of the halo/thick-disc phase and the beginning of the thin-disc phase. 
This effect is confirmed by the relation between   [Fe/O] and [O/H] observed in local stars (Chiappini et al. 1997; Gratton et al. 2000) 
and can be naturally reproduced once a SF threshold is adopted (Chiappini et al. 2001).

Moreover, in the standard model the adoption of a SF threshold causes 
numerous oscillations in the calculated SFH, which is also reflected in the SNRs.

Within the first 2 Gyr of evolution, the SFRs of the standard model and the ones calculated for the models adopting the IGIMF are 
all very similar. The SF hiatus is visible also in the models with the IGIMF, and this indicates that 
the threshold effect is important in all the four cases studied in Fig.~\ref{snr}. 
For times larger than 2 Gyr, 
the SFRs for models computed with the IGIMF are not very sensitive to the adoption of the SF threshold, with the exception 
of the model with $\beta=2.35$, which  experiences some threshold-induced star-formation  
gasping at times $> 12$ Gyr.  
The reason why the models with $\beta=1$ and $\beta=2$ are not influenced by the threshold is that in these two cases the IGIMF is 
flatter than the standard IMF (Fig. ~\ref{igimf}), 
which implies larger mass ejection rates and consequently larger gas masses available for 
star formation. 
The type Ia and type II SN rates computed with  $\beta=1$ and $\beta=2$ are higher than those 
computed with the standard IMF because the SFR values are slightly larger at any time.  
On the other hand, in the case with $\beta=2.35$  the SN rates are lower than in the standard case because 
the IGIMF is slightly steeper than the standard IMF in the mass range of the type Ia and type II SN progenitors.

The evolution of the gas surface mass density is weakly sensitive to the adopted value of $\beta$. 
As explained above, a lower  $\beta$ implies  
a higher mass ejection rate from dying stars, consequently a larger gas mass at any time. 
The fact that the stellar mass surface density computed with the standard IMF flattens at times $>10$ Gyr is basically due to the effect of the 
star formation threshold, which inhibits star formation at late evolutionary times. This flattening is not predicted for the three 
cases computed adopting the IGIMF since, as we have already seen, the star formation histories are not strongly influenced by the SF threshold. 

\begin{figure*}
\centering
\vspace{0.001cm}
\epsfig{file=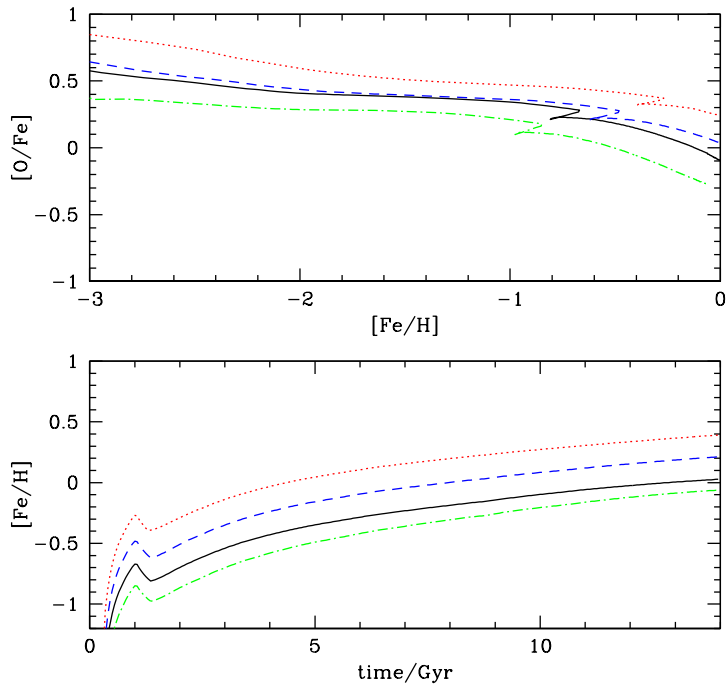,height=10cm,width=10cm}
\caption{Time evolution of the interstellar [Fe/H] (lower panel) and [O/Fe]-[Fe/H] relation (upper panel) 
computed with the 
standard IMF (solid lines) and with the IGIMF assuming $\beta=1$ (dotted lines), $\beta=2$ (dashed lines) and $\beta=2.35$ (dot-dashed lines). 
For all the models we have assumed the same value for the star formation efficiency, $\nu = 1 \, $Gyr$^{-1}$.}
\label{abu}
\end{figure*}

In Fig.~\ref{abu}, we show the predicted 
time evolution of the [Fe/H] (lower panel) and the 
[O/Fe] - [Fe/H] plot. In general, lower values of $\beta$, corresponding to flatter IGIMFs, produce at any time higher [Fe/H] values  and higher  
[$\alpha$/Fe] values at a given metallicity.  

\subsection{Local observables and their dependence on model parameters}
The observables considered in this paper, summarized in Tab. 2, are:
\begin{itemize} 
\item the abundance ratios, observed in local 
stars of various  metallicities; 
The abundance ratios  between two elements formed on different timescales 
are very useful diagnostics,  since they allow us to constrain 
the star formation history of the studied system. 
They provide us with information on the relative 
roles of various stellar sources in the chemical enrichment of the interstellar medium (Matteucci 2001). 
In particular, the study of [$\alpha$/Fe]\footnote{All the abundances between two different elements X and Y 
are expressed as $[X/Y]=log(X/Y)-log(X/Y)_{\odot}$, where  $(X/Y)$ and $(X/Y)_\odot$ are 
the ratios between the mass fractions of X and Y in the ISM and in the Sun, respectively. 
We use the set of solar abundances as determined by Grevesse at al. (2007).} is of major importance, 
owing to the difference in the timescales for $\alpha$-elements and Fe production. 
The abundance ratios are sensitive to the assumption of the IMF and to the star formation history. \\
\item the solar abundances, which are useful to test whether our models correctly reproduce the metallicity of the 
Sun at the epoch when the Solar System formed, i.e. $\sim 4.5$ Gyr ago; 
The solar abundances provide information on the integrated star formation history. 
\item the observed stellar metallicity distribution, which tells us the differential distribution of the 
living stars as a function of [Fe/H]; 
\item the present-day type Ia and type II SN rates, which are sensitive to the star formation history, to the stellar IMF and to the fraction of 
binary systems able to produce type Ia SNe; 
\item the present-day gas surface mass density, sensitive mostly to the star formation history
\item the present-day stellar surface density, depending mainly on the star formation history
\item the present-day mass function, depending on the combination of the initial mass function and the star formation history. 
\end{itemize} 
In the following, we will present all our results obtained with the IGIMF assuming three different values for the index $\beta$ of the 
star cluster mass function. 
The results obtained with the IGIMF are compared to those obtained by means of the standard IMF. 
In each case, we aim at obtaining  the best match between our models and 
the set of observational constraints by varying the star formation efficiency  
$\nu$ . 
In every single case for $\beta$, the best model is the one providing the best match simultaneously 
to the calculated SN rates, gas and stars mass surface densities,  
solar abundances, abundance ratios and present-day mass function. 

We do not consider the infall timescale as a free parameter.  
Our assumption is based on the fact that numerical dynamical models 
for disc galaxy formation indicate infall time-scales of several Gyr (Larson 1976; Samland et al. 1997). 
An infall timescale of 5-6 Gyr is indicated also by cosmological SPH simulations of disc galaxies 
in a standard  $\Lambda$-cold dark matter cosmology (Sommer-Larsen et al. 2004). 
However,   we have tested the effects of varying the infall timescale, verifying that variations of the e-folding time 
$\tau$ of 1-2 Gyr have a negligible impact on our results. 

\begin{table*}
\vspace{0cm}
\begin{flushleft}
\caption[]{Fractional mass in the Sun for various elements as observed by Grevesse et al. (2007, second column) and as predicted by means of our model by assuming 
the standard IMF (third column) and for our best models adopting the IGIMF computed with $\beta=1$ (fourth column), $\beta=2$ (fifth column), and $\beta=2.35$ (sixth column).  }
\begin{tabular}{l|lllll}
\noalign{\smallskip}
\hline
\hline
\noalign{\smallskip}
 Element             &   Obs.           & standard IMF   & IGIMF       & IGIMF       & IGIMF \\ 
                     &  (G07)           &                & $\beta=1$   & $\beta=2$   & $\beta=2.35$ \\ 
\noalign{\smallskip}
\noalign{\smallskip}
\hline
\noalign{\smallskip}
     H               &  0.7395        &    0.735         &  0.732	&   0.721        &     0.730    \\ 
     He              &  0.2485        &    0.253         &  0.251       &   0.260        &     0.259    \\ 
     C               &  2.18(-3)      &    1.83(-3)      &  1.72(-3)    &   2.57(-3)     &      2.47(-3) \\ 
     O               &  5.41(-3)      &    5.5(-3)       &  8.819(-3)   &   8.20(-3)     &      3.35(-3) \\ 
     Mg              &  6.01(-4)      &    7.0(-4)       &  8.89(-4)    &   1.10(-3)     &      4.94(-4) \\ 
     Si              &  6.70(-4)      &    7.6(-4)       &  9.19(-4)    &   1.20(-3)     &      6.35(-4) \\ 
     Fe              &  1.17(-3)      &    1.21(-3)      &  1.19(-3)    &   1.96(-3)     &      1.32(-3)  \\ 
      Z              &  0.012         &    0.0126        &  0.016       &   0.019        &      0.010 \\ 
\hline
\hline
\end{tabular}
\label{solar_abu}
\end{flushleft}
\end{table*}

\begin{figure*}
\centering
\vspace{0.001cm}
\epsfig{file=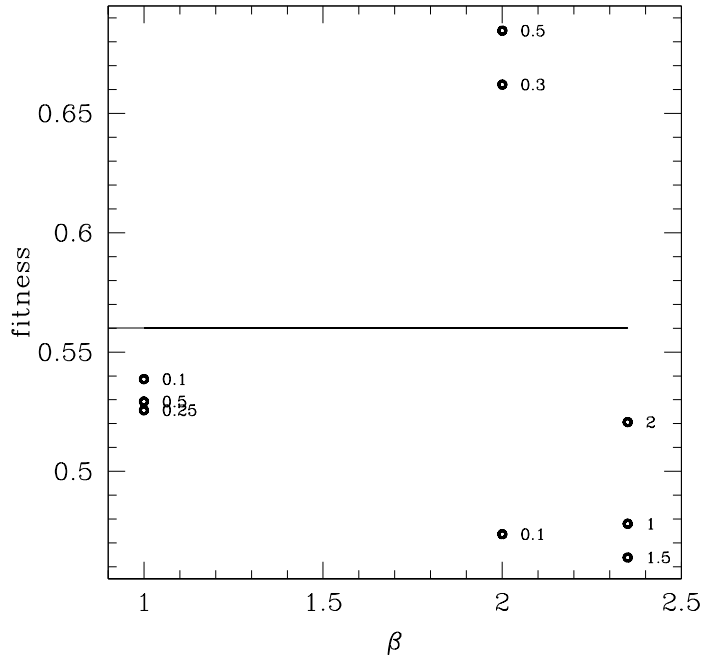,height=10cm,width=10cm}
\caption{Open circles: fitness (defined in Section~\ref{results}) as a function of $\beta$ 
for various models, each one characterized by a particular 
star formation efficiency $\nu$, indicated by the number beside each open circle. The horizontal line indicates 
the fitness value computed for the standard model. 
From this figure it is clear that the model with $\beta=2$ and $\nu=0.5$ is the one providing the best fit to the 
set of observables studied in this paper. }
\label{fit}
\end{figure*}

\begin{figure*}
\centering
\vspace{0.001cm}
\epsfig{file=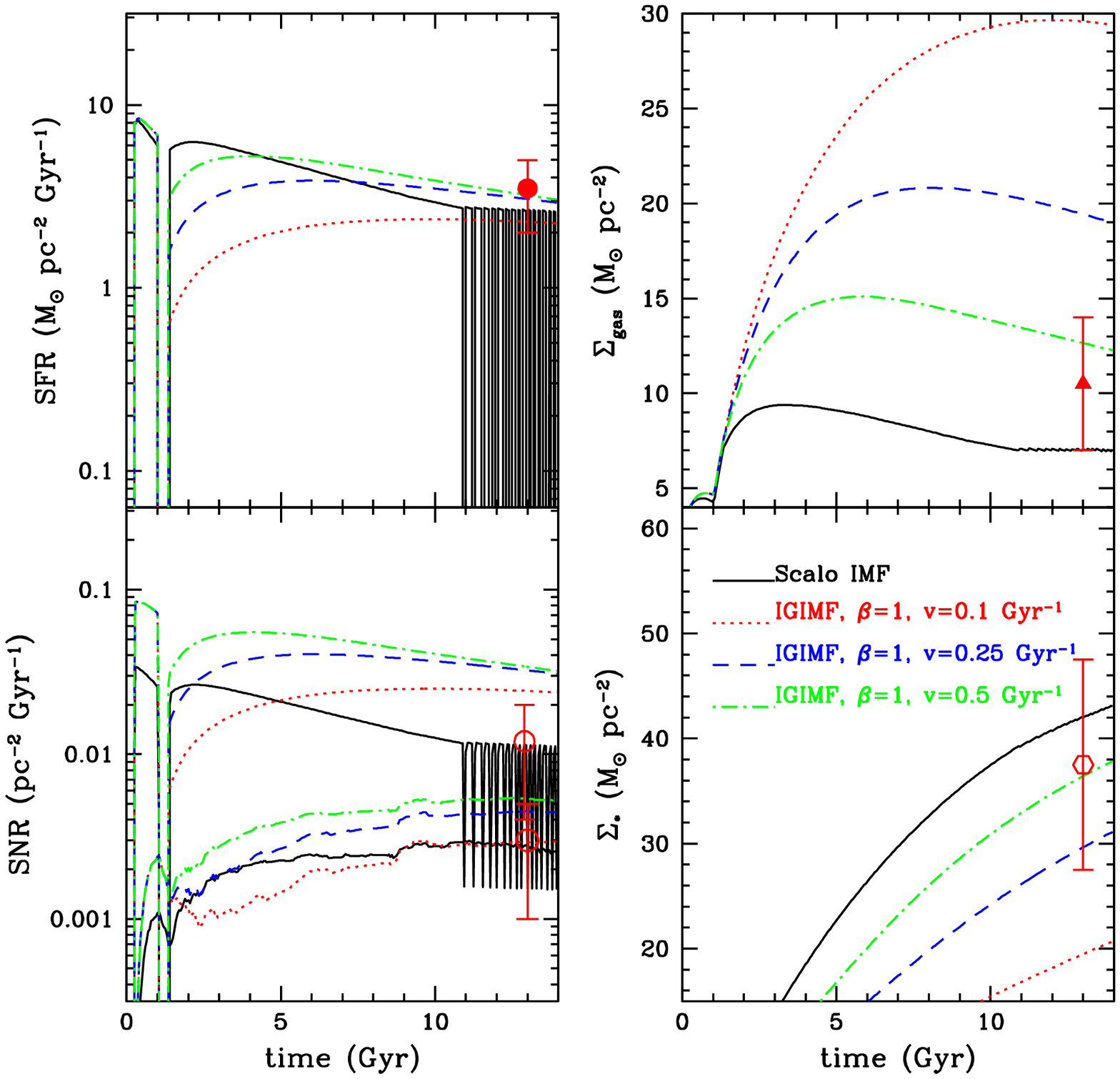,height=10cm,width=10cm}
\caption{From top-left, clockwise: 
calculated time evolution of the star formation history, of the gas surface density, 
stellar surface density and SN rates computed by means of the solar neighbourhood model with the standard IMF 
(solid lines) and by means of three models with the IGIMF, in the case $\beta=1$  and assuming three different SF efficiencies: 
$\nu=0.1$ Gyr$^{-1}$ (dotted lines), $\nu=0.25$ Gyr$^{-1}$ (dashed lines), 
and $\nu=0.5$ Gyr$^{-1}$ (dash-dotted lines). 
In the panel showing the SN rate evolution, the lower  and upper curves represent the calculated type Ia and
 type II SN rates, respectively. Observational data are reported in Tab.~\ref{data}. 
}
\label{snr_beta1}
\end{figure*}

\begin{figure*}
\centering
\vspace{0.001cm}
\epsfig{file=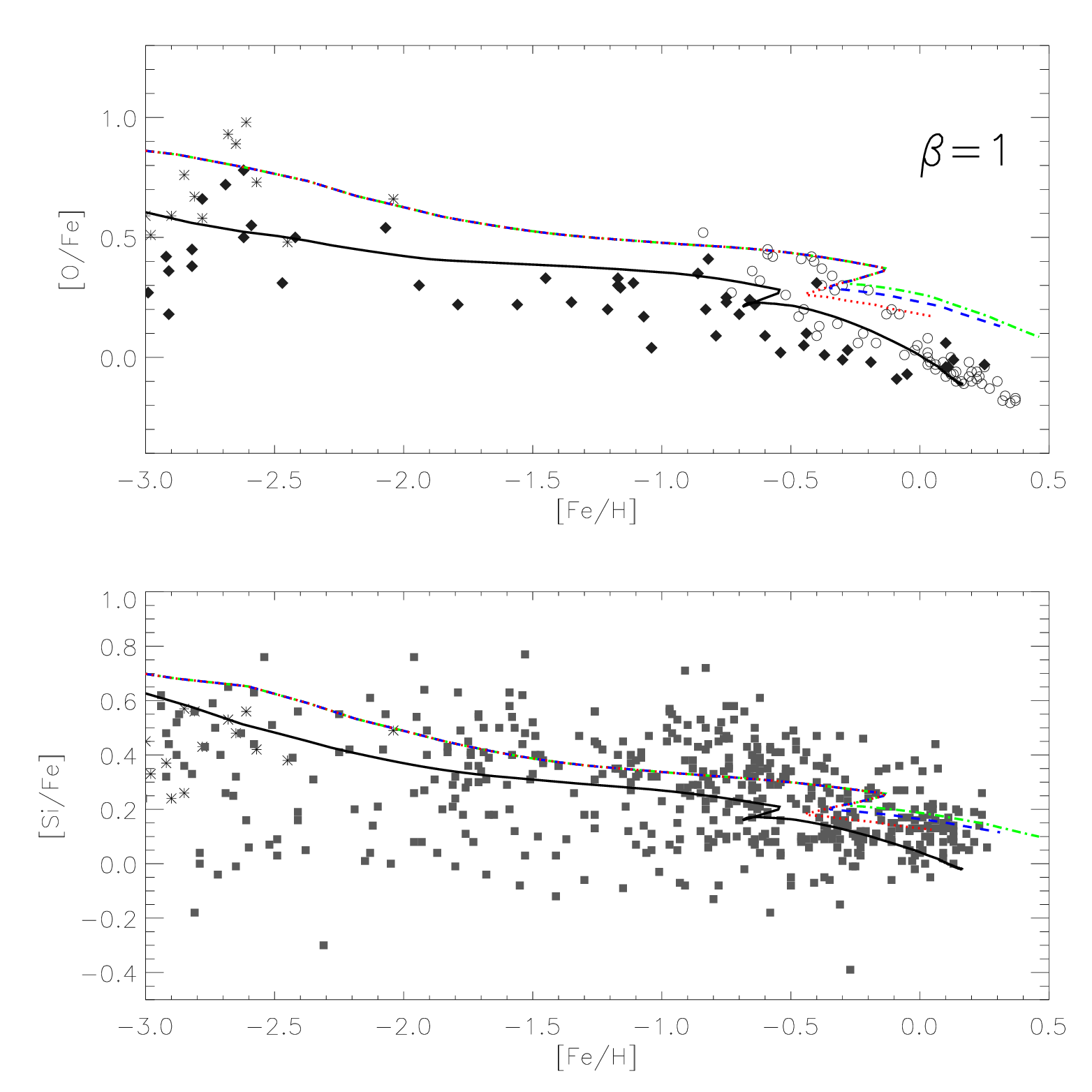,height=10cm,width=10cm}
\caption{Predicted [O/Fe]-[Fe/H] (upper panel) and [Si/Fe]-[Fe/H] (lower panel) 
computed by means of the solar neighbourhood model with the standard IMF 
(solid lines) and by means of three models 
with the IGIMF, in the case $\beta=1$ 
and assuming three different SF efficiencies: 
$\nu=0.1$ Gyr$^{-1}$ (dotted lines), $\nu=0.25$ Gyr$^{-1}$ (dashed lines), 
and $\nu=0.5$ Gyr$^{-1}$ (dash-dotted lines). 
The predictions are compared 
to observational data from various authors. Asterisks: Cayrel et al. (2004); open circles: Bensby et al. (2003); 
solid diamonds: Fran\c cois et al. (2004). The solid squares 
are from a compilation of data by Cescutti (2008).}
\label{elfe_beta1}
\end{figure*}

\begin{figure*}
\centering
\vspace{0.001cm}
\epsfig{file=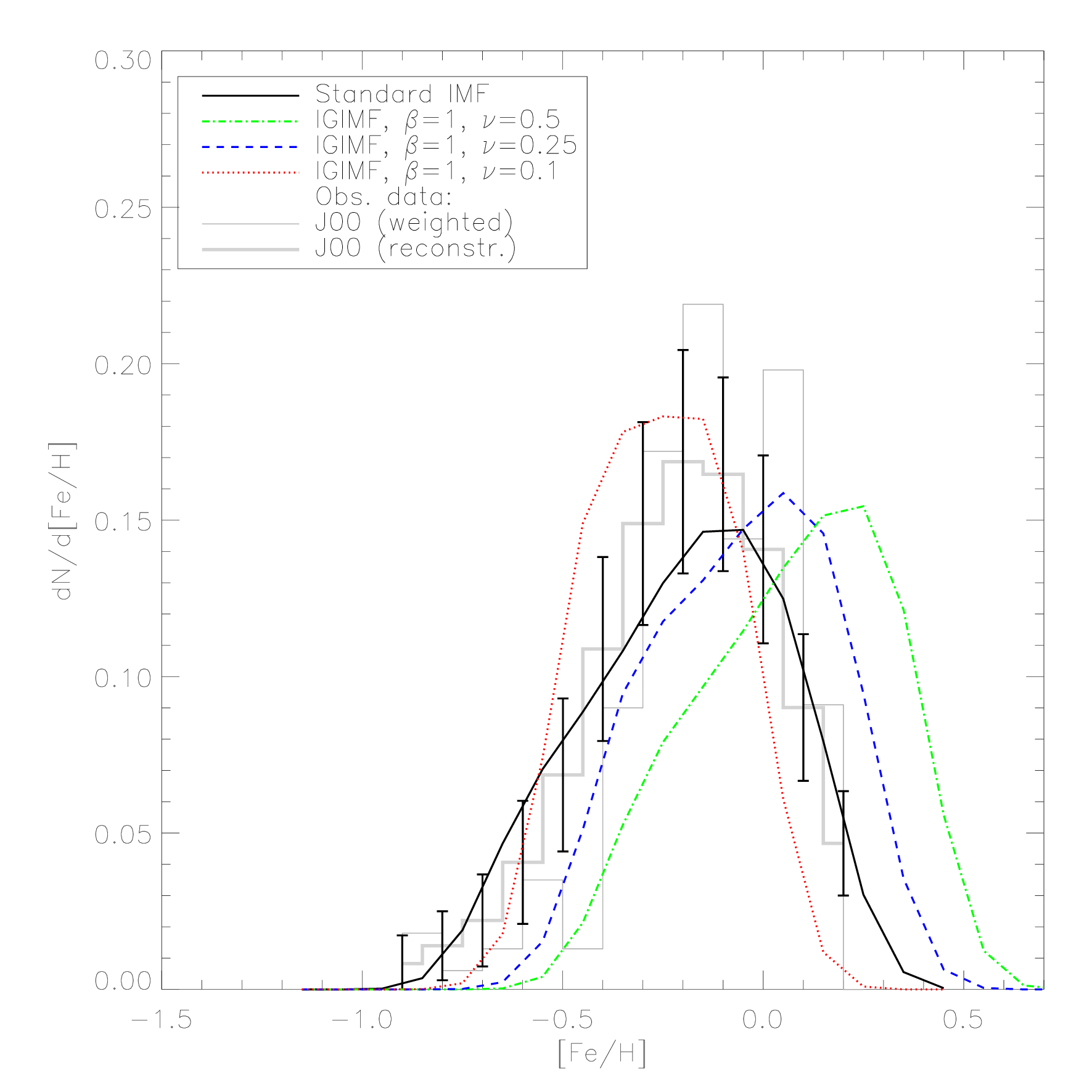,height=10cm,width=10cm}
\caption{Present-day stellar metallicity distribution in the Solar Neighbourhood. 
The solid line, the dotted line, the dashed line and the dash-dotted line represent the predicted stellar metallicity 
distribution  computed with the standard IMF and 
with the IGIMF in the case $\beta=1$ and  SF efficiency $\nu=0.1$ Gyr$^{-1}$, 
$\nu=0.25$ Gyr$^{-1}$, 
 and  $\nu=0.5$ Gyr$^{-1}$, respectively. 
The thin and thick solid histograms represent the observational 
weighted and reconstructed SMD from Jorgensen (2000), respectively. For the reconstructed distribution, error bars are plotted. }
\label{dndfe_beta1}
\end{figure*}

\subsection{$\beta=1$}
\label{beta1}
In Fig.~\ref{snr_beta1}, we show 
the star formation history, the time evolution of the type Ia and type II SN rates, of the gas surface density 
and of the stellar mass density for three models with $\beta=1$ and assuming various SF efficiencies $\nu$, 
compared to the results obtained with the standard IMF. 
By assuming $\beta=1$ for the embedded cluster mass function, the model which  best reproduces our set of 
observational constraints is  characterized by a SF efficiency $\nu=0.1$ Gyr$^{-1}$.  
This can be seen from Fig.~\ref{fit}, where we show how the ``fitness'' quantity defined in section~\ref{results} behaves as 
a function of $\beta$. Among the models with $\beta=1$, the one with the highest fitness value is that 
with $\nu=0.1$ Gyr$^{-1}$.

From Fig.~\ref{snr_beta1} it is clear that there are some observables which cannot be reproduced accurately, 
such as the present 
gas and stellar mass densities. 
Furthermore, 
All the models with $\beta=1$ tend to overestimate the type II SN rate. 
This result should be expected, since the assumption $\beta=1$ 
produces an IMF flatter than the standard one, hence richer in massive stars. 

In Table 3, we present the predicted solar abundances for the most important heavy elements, 
obtained with the standard IMF and with the best models 
using  the IGIMFs for various $\beta$. We limit our calculations to the cases of the most important chemical elements, 
for which the standard model 
reproduces the observed abundance pattern with the most accurate precision.

In the case $\beta=1$, the IGIMF being flatter than the standard IMF implies heavy element abundances higher than the solar ones. 
By decreasing further the SF efficiency, it is possible to improve the match between the observed solar 
abundances and the predicted ones, but at the expense of the match of the local gas and star surface mass densities. 

In Fig.~\ref{elfe_beta1}, we present the abundance ratios as a function of metallicity, traced by $[Fe/H]$, 
computed by means of our standard chemical evolution model 
and compared to three models with $\beta=1$ and different SF efficiencies.  
As expected, owing to the excess of massive stars and to the very high type II SN rates, 
the predictions obtained with the IGIMF and $\beta=1$ overestimate all the [$\alpha$/Fe] 
ratios. It is also interesting to note  that the results are weakly dependent 
on the SF efficiencies for values in the range $0.1 \le \nu/$Gyr$^{-1} \le 0.5$.\\
In Fig.~\ref{dndfe_beta1}, we show instead the stellar metallicity distribution computed by means of our 
standard IMF and with the IGIMF, showing results for three different SF efficiencies.  
The standard IMF allows us to reproduce the observed stellar metallicity distribution (SMD) with good accuracy, 
concernig either the low-metallicity tail and the peak metallicity value.
In Fig.~\ref{dndfe_beta1}, we show both the weighted and reconstructed SMDs as observed by Jorgensen (2000). 
The error bars are provided only for the reconstructed SMD and are also plotted in Fig. ~\ref{dndfe_beta1}. 
By means of our best model  computed with the IGIMF and $\beta=1$, the predicted SMD is not well 
reproduced. The peak metallicity is underestimated, and the number of stars with metallicity in the range 
-0.5$\le$[Fe/H]$<-0.25$ is overestimated. On the other hand, the high-metallicty tail is remarkably 
in disagreement with the observations. Models characterised by higher star formation efficiencies lead to an overabundance of 
stars with metallicity [Fe/H]$>$0 with respect to the observations.\\
We can conclude that, by assuming $\beta=1$, no model can satisfactorily reproduce at the same time 
all the observables considered in this work.

\begin{figure*}
\centering
\vspace{0.001cm}
\epsfig{file=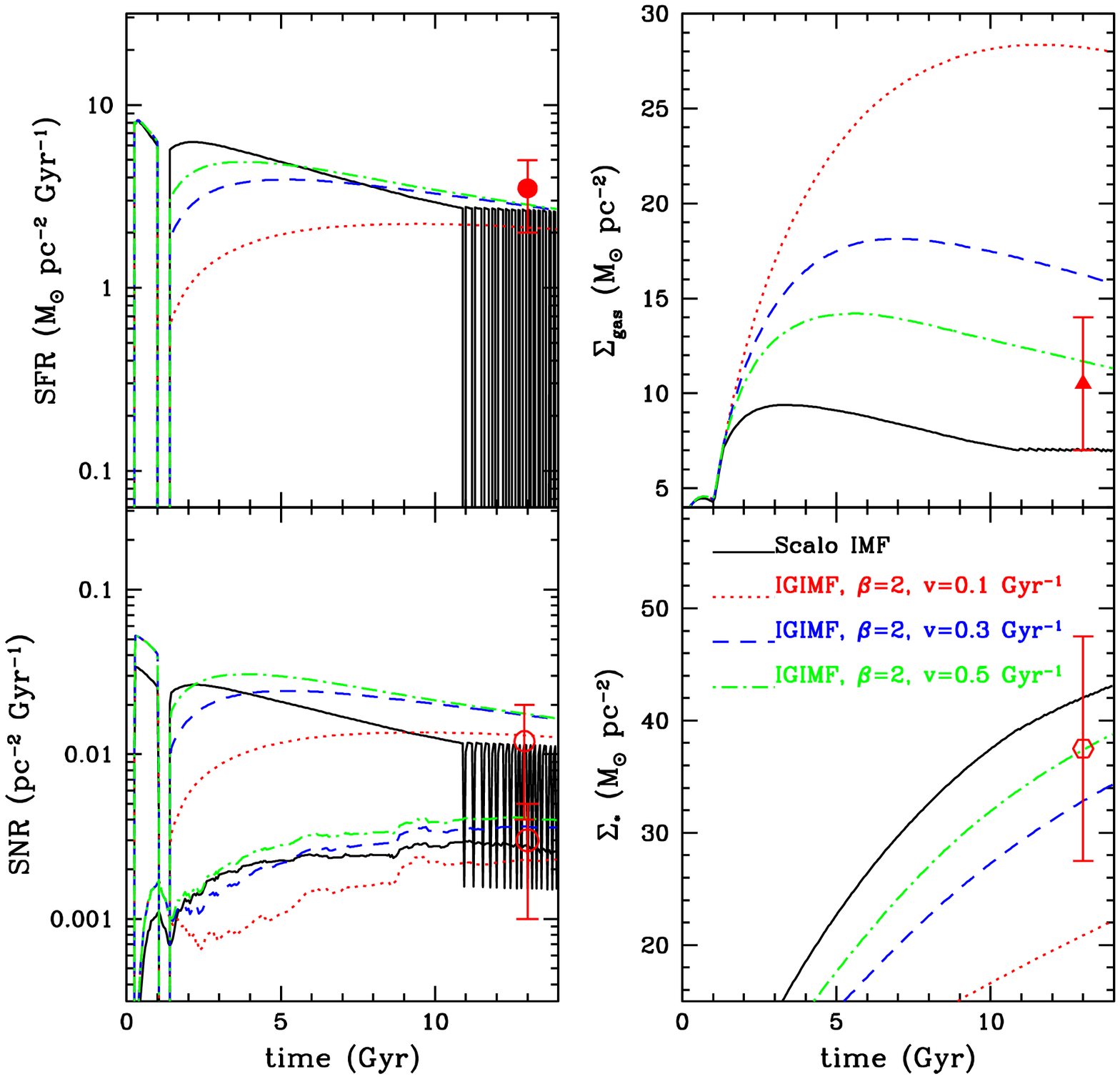,height=10cm,width=10cm}
\caption{From top-left, clockwise: 
predicted time evolution of the star formation history, of the gas surface density, 
stellar surface density and SN rates computed by means of the solar neighbourhhod model with the standard IMF 
(solid lines) and by means of three models with the IGIMF, in the case $\beta=2$  and assuming three different SF efficiencies: 
$\nu=0.1$ Gyr$^{-1}$ (dotted lines), $\nu=0.3$ Gyr$^{-1}$ (dashed lines), 
and $\nu=0.5$ Gyr$^{-1}$ (dash-dotted lines). 
Symbols as in Fig.~\ref{snr_beta1}.}
\label{snr_beta2}
\end{figure*}

\begin{figure*}
\centering
\vspace{0.001cm}
\epsfig{file=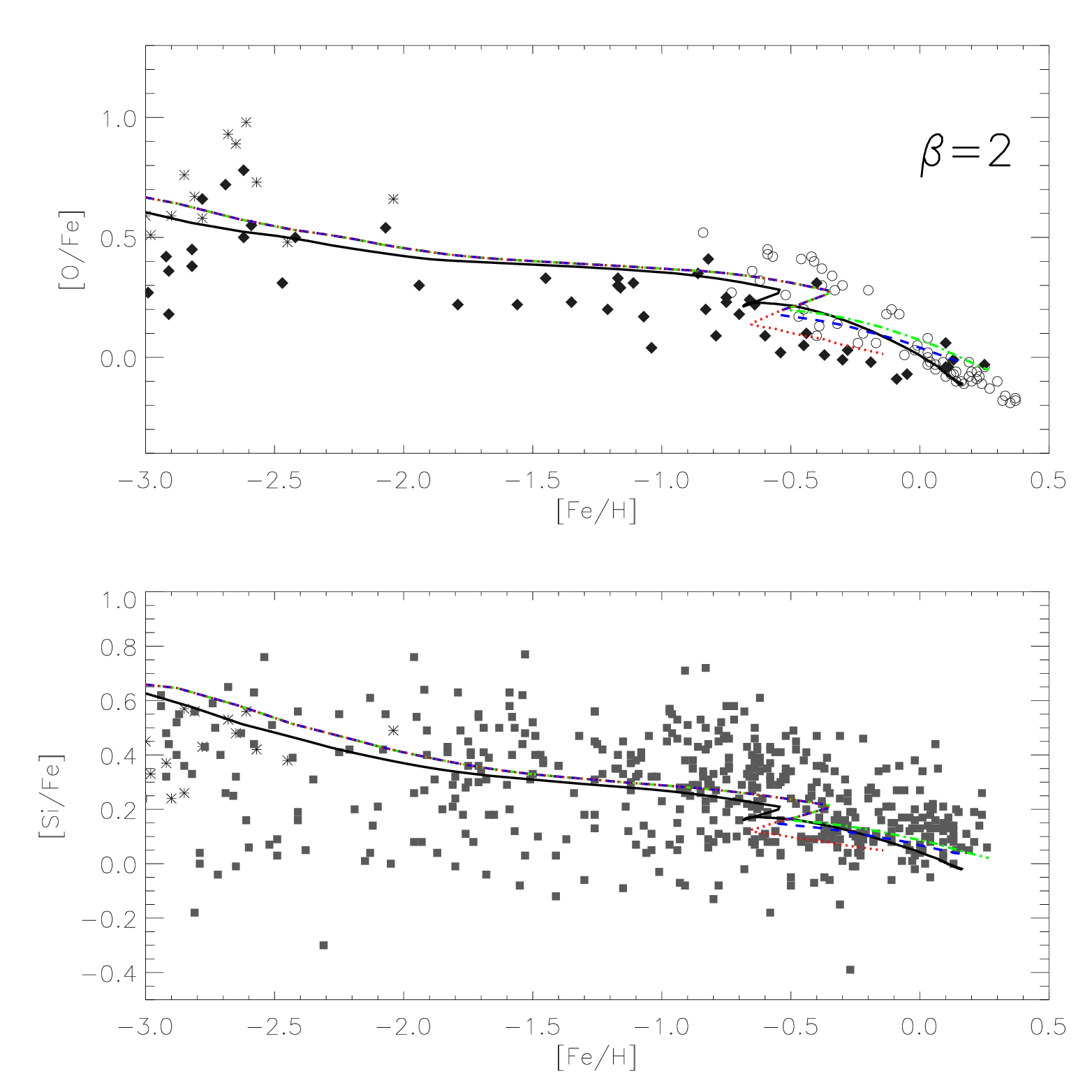,height=10cm,width=10cm}
\caption{Predicted [O/Fe]-[Fe/H] (upper panel) and  [Si/Fe]-[Fe/H] (lower panel) 
computed by means of the solar neighbourhood model with the standard IMF 
(solid lines) and by means of three models 
with the IGIMF, in the case $\beta=2$ 
and assuming three different SF efficiencies: 
$\nu=0.1$ Gyr$^{-1}$ (dotted lines), $\nu=0.3$ Gyr$^{-1}$ (dashed lines), 
and $\nu=0.5$ Gyr$^{-1}$ (dash-dotted lines), 
compared to observational data from various authors (see caption of Fig.~\ref{elfe_beta1}). }
\label{elfe_beta2}
\end{figure*}

\begin{figure*}
\centering
\vspace{0.001cm}
\epsfig{file=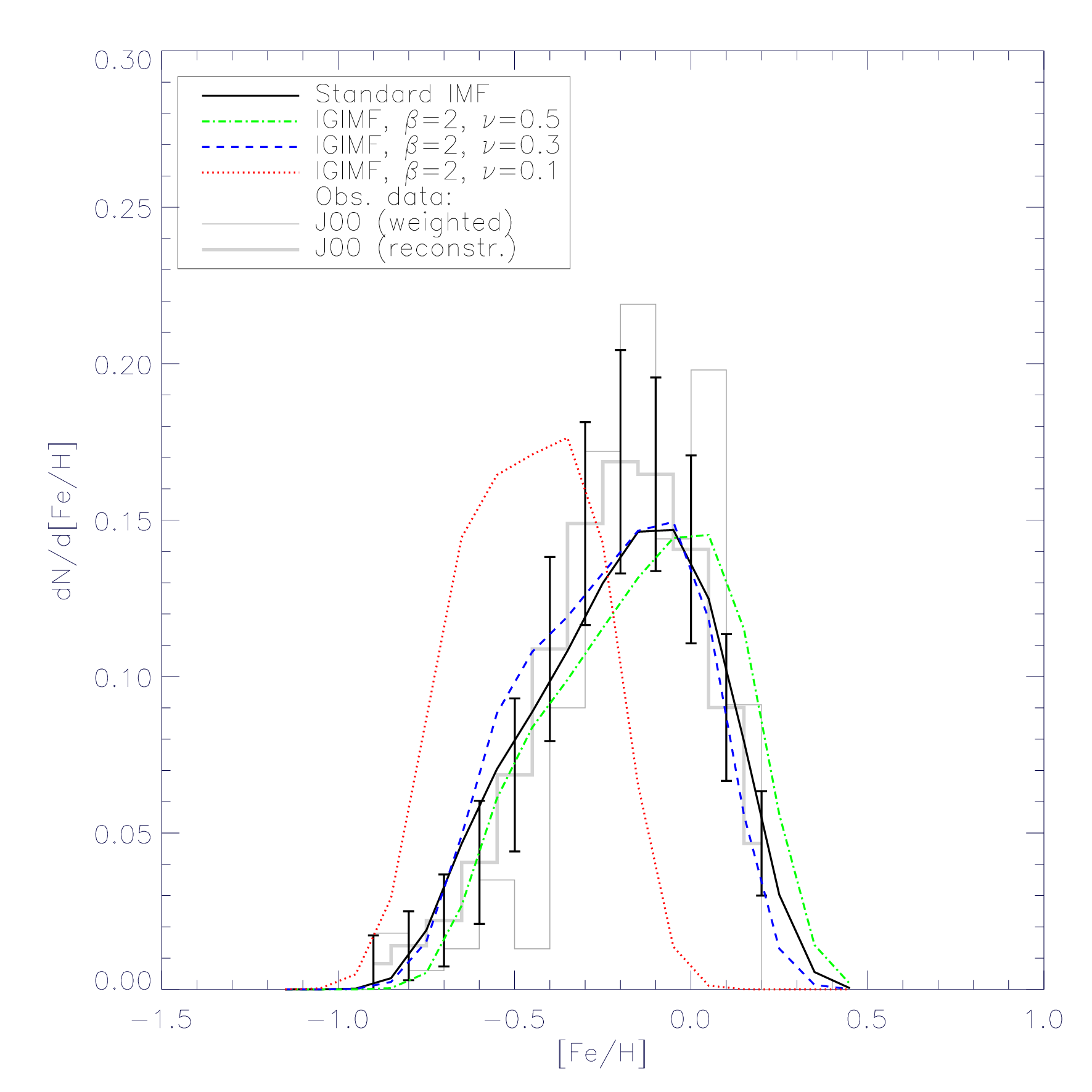,height=10cm,width=10cm}
\caption{Present-day stellar metallicity distribution in the Solar Neighbourhood. 
The solid line, the dotted line, the dashed line and the dash-dotted line represent the predicted SMDs computed with the 
standard IMF, 
with the IGIMF in the case $\beta=2$ and  SF efficiency $\nu=0.1$ Gyr$^{-1}$, 
$\nu=0.3$ Gyr$^{-1}$, 
and $\nu=0.5$ Gyr$^{-1}$, respectively. 
The histograms are the observational data described in Fig. \ref{dndfe_beta1}. 
}
\label{dndfe_beta2}
\end{figure*}

\subsection{$\beta=2$}
\label{beta2}
In Fig.~\ref{snr_beta2}, we present the star formation history, the time evolution of the type Ia and type II SN rates, 
of the gas surface density 
and of the stellar mass density for three models with $\beta=2$ and different SF efficiencies $\nu$, 
compared to the results with the standard IMF and to the available 
local observations. 
The case with $\beta=2$ is more promising than the previous one for reproducing the constraints considered in this work. 
In this case,  
the model which  best reproduces our set of 
observational constraints is  characterized by a SF efficiency $\nu=0.5$ Gyr$^{-1}$ (see Fig.~\ref{fit}), althogh  
also the model with  SF efficiency $\nu=0.3$ Gyr$^{-1}$ provides satisfactory results, as shown by the fitness test. 
We have also tested models with SF efficiency $\nu>0.5$ Gyr$^{-1}$, finding fitness values lower than the one found with $\nu=0.3-0.5$ Gyr$^{-1}$.

Both type Ia and II SNRs are larger than those predicted with the standard IMF. However, both the present-day SN 
rates predicted with the IGIMF are consistent with the local observations by Cappellaro (1996). The 
gas and stellar surface densities  predicted with the best model with the IGIMF are 
in agreement with the observed values.

In the fourth column of Table 3, we present the solar abundances predicted with the best model with the IGIMF and $\beta=2$. 
As can be seen from Fig.~\ref{snr_beta2} and  Table 3, a satisfactory 
match between the predictions and the observations is achievable in this case, although the heavy element abundances are slightly overestimated.

In Fig.\ref{elfe_beta2} we show the abundance ratios as a function of [Fe/H], compared to available observations in local stars. 
In this case, the main trends of the observed abundance ratios as a function of [Fe/H] are well accounted for, as well as 
the abundance ratios at the solar [Fe/H].

The results  obtained with this choice of $\beta$ are quite similar to the ones achieved with the standard IMF. 
This result could be expected since, as shown in Fig.~\ref{igimf}, 
in the intermediate case with $\beta=2$  the IGIMF is very similar to the standard IMF.  
However, the IGIMF is slightly more top-heavy than the standard IMF. 
For the best model, this translates into [$\alpha/Fe$]  
ratios sligthly larger than those computed with the standad IMF. 

As shown in Fig.~\ref{dndfe_beta2},  
it is possible to obtain good results also for the SMD, with very little difference between 
the model characterized by $\nu=0.3$ Gyr$^{-1}$ and $\beta=2$ and the standard model.
 On the other hand, the model with the IGIMF $\nu=0.5$ Gyr$^{-1}$ is slightly overestimating  the number of high metalicity stars.  
However, globally, with $\beta=2$ satisfactory results can be achieved also concerning the study of the stellar 
metallicity distribution.

We can conclude that the assumption of $\beta=2$ allows us to satisfactorily reproduce the set of observational constraints considered in this work. 
This is an important result, given the fact 
that the 
IGIMF is computed from first principles.
Furthermore, this result allows us to constrain the embedded cluster mass function. In the case this function  may be 
approximated by a power law, here we have shown that 
the index  $\beta=2$ is to be favoured with respect to the case $\beta=1$. \\

The results with $\beta=2$ are very similar to those obtained with the standard IMF, 
and in principle, on the basis of the results described in this 
section, it may be difficut to discriminate between the two scenarios. 
In Sect.~\ref{PDMF}, we will suggest the use of a  diagnostic allowing us  to 
put further constraints on the IMF in the solar neighbourhood and to disentangle between the standard IMF and the IGIMF.

\begin{figure*}
\centering
\vspace{0.001cm}
\epsfig{file=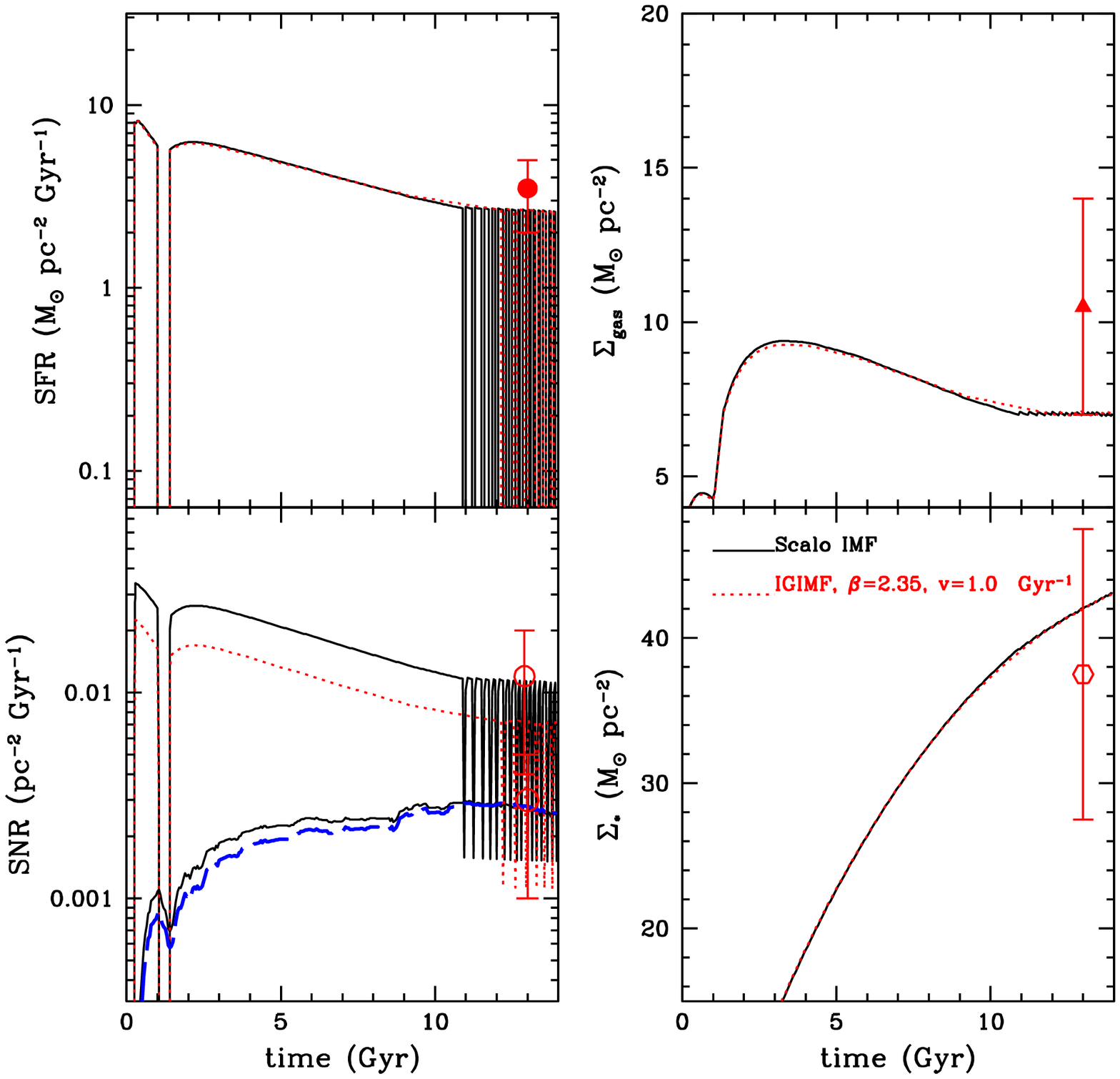,height=10cm,width=10cm}
\caption{From top-left, clockwise: 
predicted time evolution of the star formation history, of the gas surface density, 
stellar surface density and SN rates computed by means of the solar neighbourhhod model with the standard IMF 
(solid lines) and by means of the best model with the IGIMF, in the case $\beta=2.35$ and a SF efficiency 
$\nu=1$ Gyr$^{-1}$  (dotted lines). 
In the panel showing the SN rate evolution, the lower  and upper curves represent the predicted type Ia and the 
predicted type II SN rates, respectively. Symbols as in Fig.~\ref{snr_beta1}.}
\label{snr_beta2_35}
\end{figure*}

\begin{figure*}
\centering
\vspace{0.001cm}
\epsfig{file=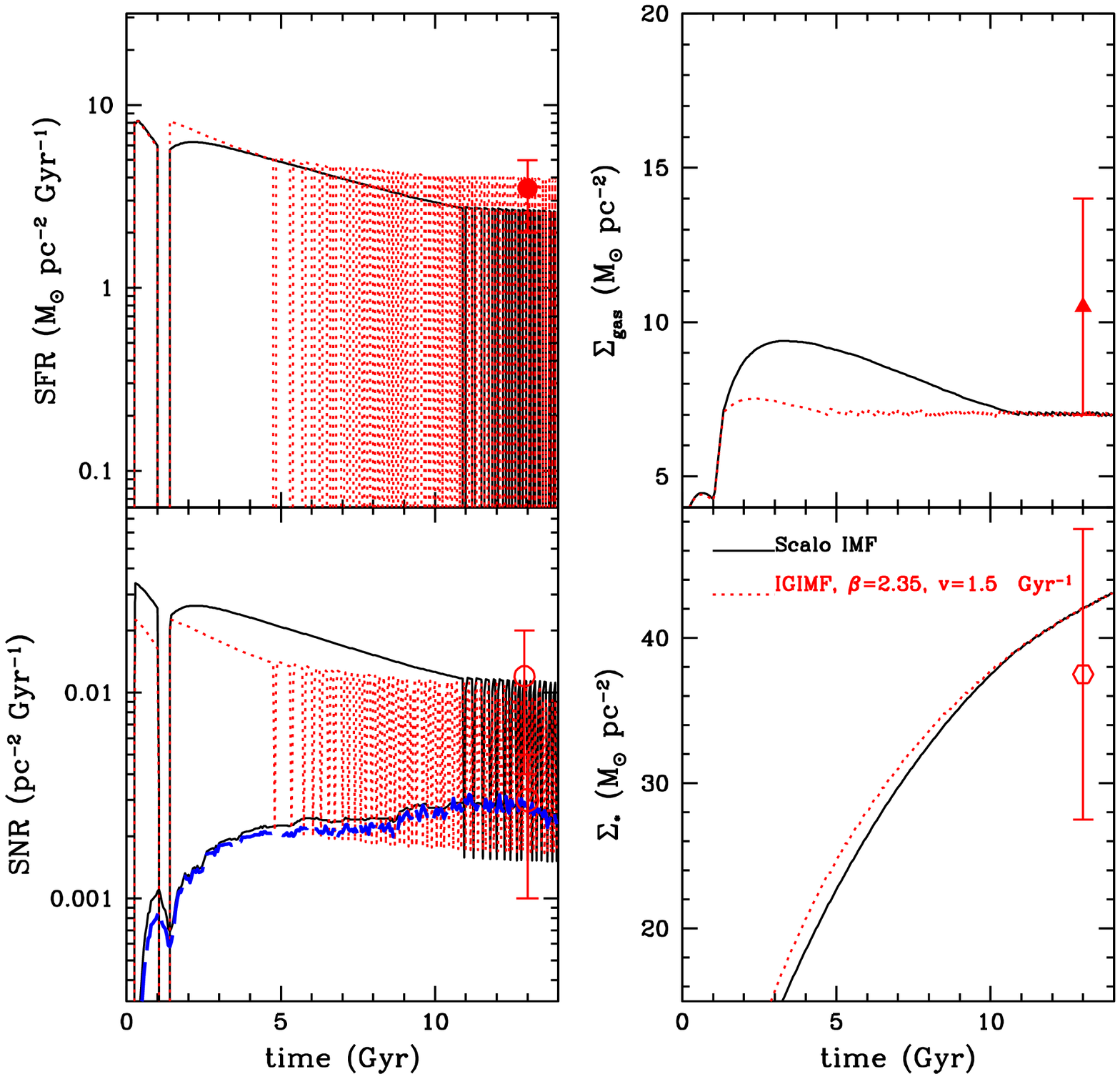,height=10cm,width=10cm}
\caption{From top-left, clockwise: 
predicted time evolution of the star formation history, of the gas surface density, 
stellar surface density and SN rates computed by means of the solar neighbourhhod model with the standard IMF 
(solid lines) and by means of the best model with the IGIMF, in the case $\beta=2.35$ and a SF efficiency 
$\nu=1.5$ Gyr$^{-1}$  (dotted lines)
In the panel showing the SN rate evolution, the lower  and upper curves represent the predicted type Ia and the 
predicted type II SN rates, respectively. Symbols as in Fig.~\ref{snr_beta1}.}
\label{snr_beta2_35_v15}
\end{figure*}

\begin{figure*}
\centering
\vspace{0.001cm}
\epsfig{file=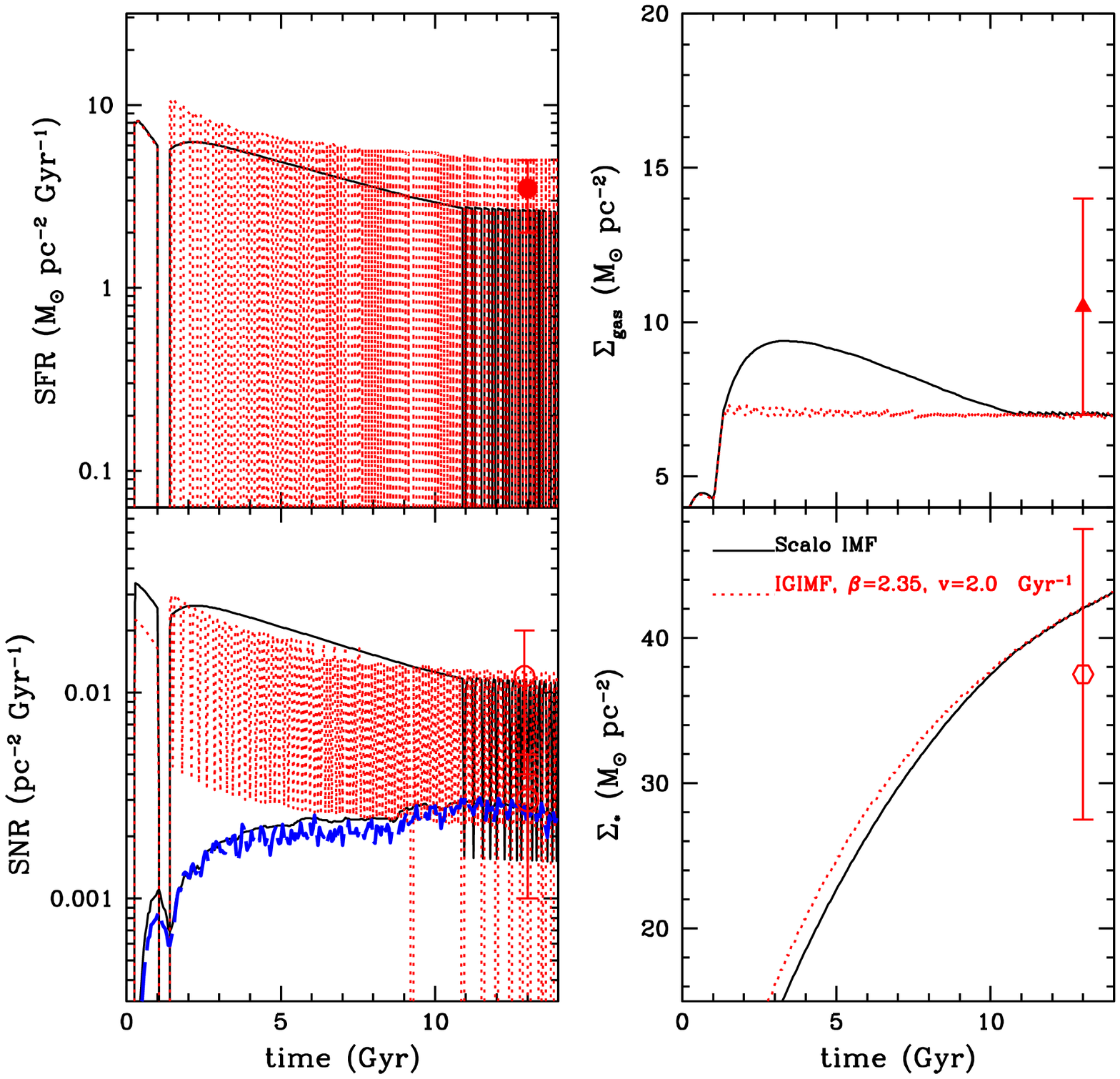,height=10cm,width=10cm}
\caption{From top-left, clockwise: 
predicted time evolution of the star formation history, of the gas surface density, 
stellar surface density and SN rates computed by means of the solar neighbourhhod model with the standard IMF 
(solid lines) and by means of the best model with the IGIMF, in the case $\beta=2.35$ and a SF efficiency 
$\nu=2$ Gyr$^{-1}$ (dotted lines). 
In the panel showing the SN rate evolution, the lower  and upper curves represent the calculated type Ia and 
type II SN rates, respectively. Symbols as in Fig.~\ref{snr_beta1}. }
\label{snr_beta2_35_v2}
\end{figure*}

\begin{figure*}
\centering
\vspace{0.001cm}
\epsfig{file=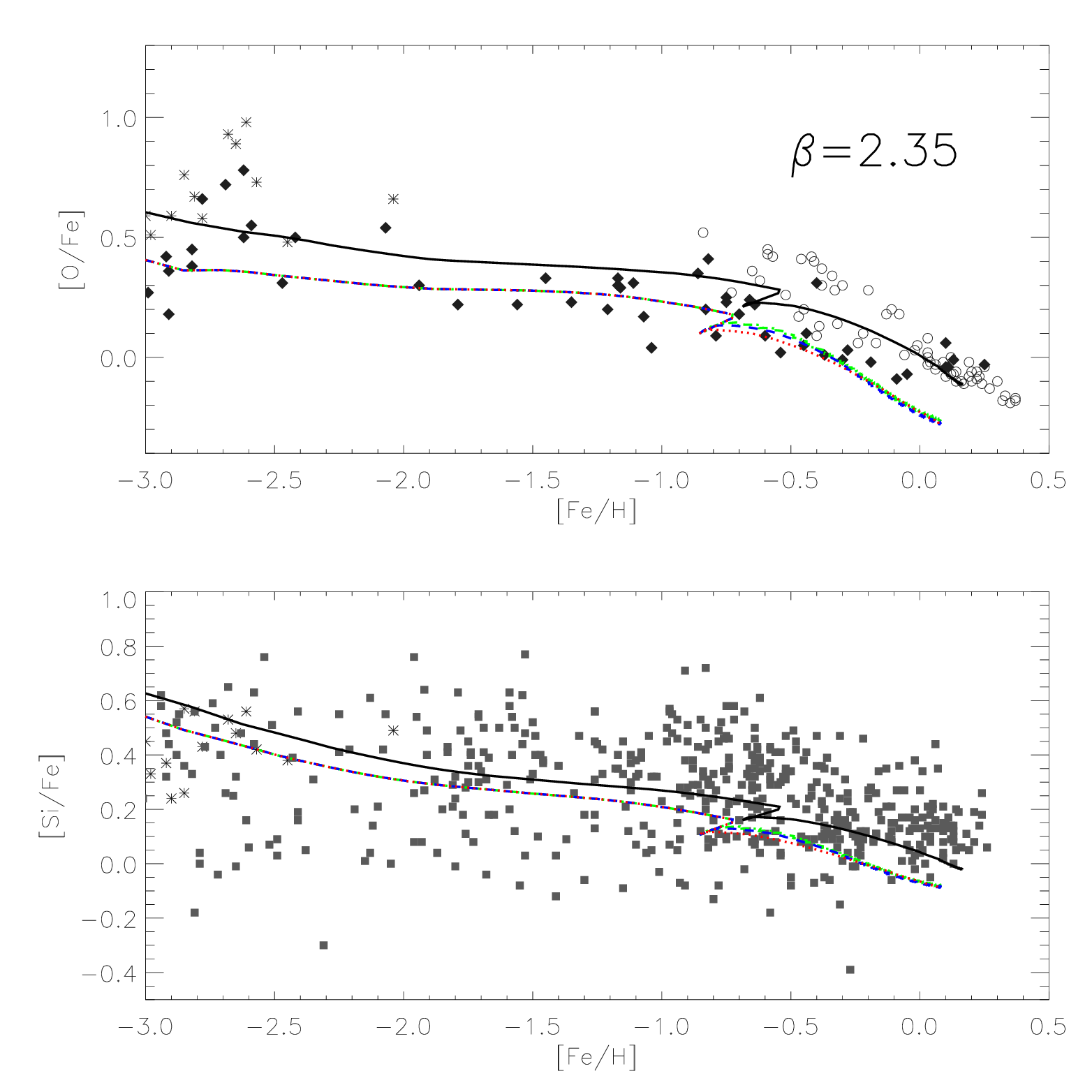,height=10cm,width=10cm}
\caption{Predicted [O/Fe]-[Fe/H] (upper panel) and [Si/Fe]-[Fe/H] (lower panel) 
computed by means of the solar neighbourhood model with the standard IMF 
(solid lines) and by means of three models 
with the IGIMF, in the case $\beta=2.35$ 
and assuming three different SF efficiencies: 
$\nu=1$ Gyr$^{-1}$ (dotted lines), $\nu=1.5$ Gyr$^{-1}$ (dashed lines), 
and $\nu=2$ Gyr$^{-1}$ (dash-dotted lines), 
compared to observational data from various authors (see caption of Fig.~\ref{elfe_beta1}). }
\label{elfe_beta2_35}
\end{figure*}

\begin{figure*}
\centering
\vspace{0.001cm}
\epsfig{file=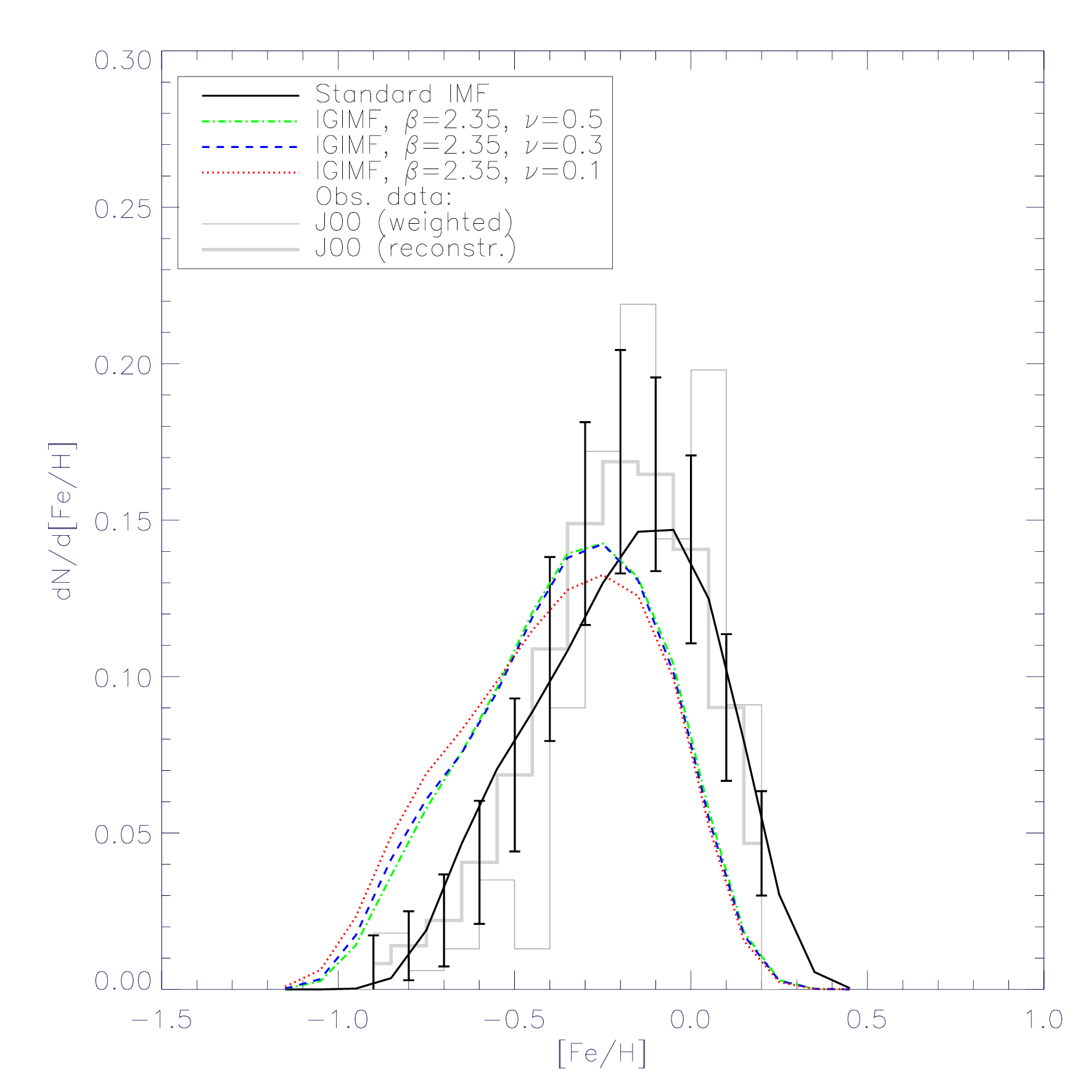,height=10cm,width=10cm}
\caption{Present-day stellar metallicity distribution in the Solar Neighbourhood. 
The solid line, the dotted line, the dashed line and the dash-dotted line represent the predicted SMDs computed with the standard IMF, 
with the IGIMF (in the case $\beta=2.35$) and  SF efficiency $\nu=1$ Gyr$^{-1}$, 
with the IGIMF and  SF efficiency $\nu=1.5$ Gyr$^{-1}$, 
and with the IGIMF and  SF efficiency $\nu=2$ Gyr$^{-1}$, respectively. 
The histograms are the observational data described in Fig. \ref{dndfe_beta1}. }
\label{dndfe_beta2_35}
\end{figure*}

\subsection{$\beta=2.35$}
The assumption of  $\beta=2.35$ produces fewer  massive stars than in the standard 
case, and this has some impact on the predicted SN rates and gas and stellar surface mass densities. 
In Fig.s~\ref{snr_beta2_35}, ~\ref{snr_beta2_35_v15}, and ~\ref{snr_beta2_35_v2} 
we show the predicted time evolution of the SFR, SN rates, gas and stellar mass densities computed 
by assuming the IGIMF with $\beta=2.35$ and SF efficiencies $\nu=1$ Gyr$^{-1}$, $\nu=1.5$ Gyr$^{-1}$, and 
$\nu=2$ Gyr$^{-1}$, respectively, compared to the standard model and to the local observational values. 
In the sixth column of Table 3 we show the  solar abundances predicted with the best model with $\beta=2.35$, 
characterized by a SF efficiency 
$\nu=2$ Gyr$^{-1}$ (see Fig.~\ref{fit}). 
This implies a quicker gas consumption timescale and stronger effects of the SF threshold than in the standard case, which has a 
SF efficiency  $\nu=1$ Gyr$^{-1}$.  
While the model with the standard IMF is influenced by the SF threshold only at times $>11$ Gyr,  
the adoption of a higher SF efficiency causes 
strong threshold effects already after 1 Gyr of evolution. This is visible mainly in the predicted SF history, type II SNR and gas surface density evolution. 

For some elemnents (O,Mg), the model reproducing best the data of Fig.~\ref{snr_beta2_35} 
provides solar abundances lower than the observed ones (see Tab.~\ref{solar_abu}). 
On the other hand, as shown by Fig.~\ref{elfe_beta2_35}, the analysis of the abundance ratios as a function of 
metallicity indicates [$\alpha$/Fe] values lower than the ones computed with the standard IMF.  
Finally, in Fig. ~\ref{dndfe_beta2_35}, we show the SMD computed with the IGIMF assuming  $\beta=2.35$ compared to 
the observations and the results of the standard IMF. 
In this case, the peak metallicity computed with the IGIMF is shifted leftwards by 0.2 dex with respect to the observations and to the standard model. 
From the results discussed in this section we can conclude that by assuming $\beta=2.35$ it is not possible to reproduce at the same time all the observational 
constraints considered in this paper.

\begin{figure*}
\centering
\vspace{0.001cm}
\epsfig{file=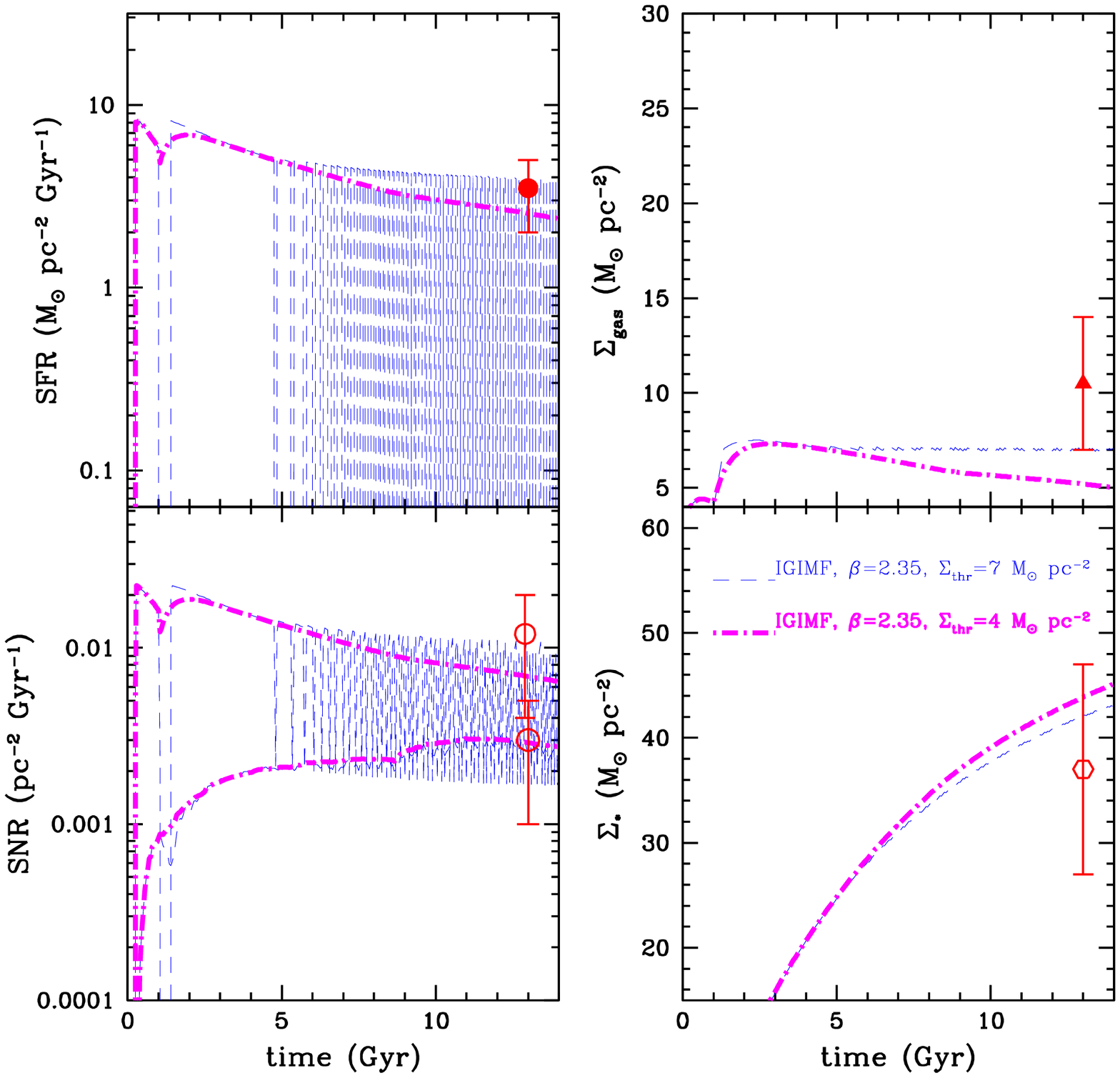,height=10cm,width=10cm}
\caption{From top-left, clockwise: 
predicted time evolution of the star formation history, of the gas surface density, 
stellar surface density and SN rates computed by means of the model with 
$\beta=2.35$ and a SF efficiency $\nu=1.5$ and adopting a SF threshold $\Sigma_{thr}=7 M_{\odot} pc^{-2}$ (thin dashed lines) 
and $\Sigma_{thr}=4 M_{\odot} pc^{-2}$ (thick dot-dashed lines). }
\label{snr_beta2_35_thr}
\end{figure*}

\begin{figure*}
\centering
\vspace{0.001cm}
\epsfig{file=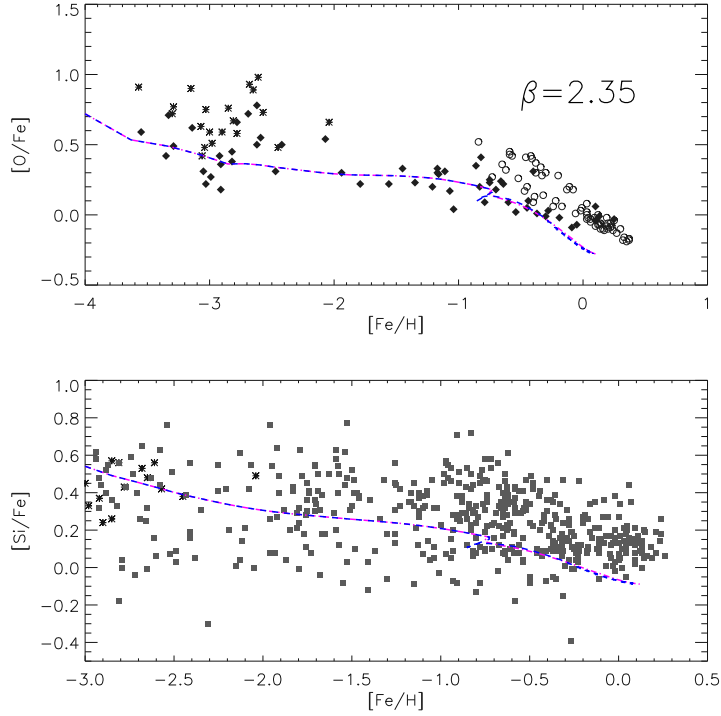,height=10cm,width=10cm}
\caption{ 
Predicted [O/Fe]-[Fe/H] (upper panel) and [Si/Fe]-[Fe/H] (lower panel) computed by means of 
the model with 
$\beta=2.35$ and a SF efficiency $\nu=1.5$ and adopting a SF threshold $\Sigma_{thr}=7 M_{\odot} pc^{-2}$ (dashed lines) 
and $\Sigma_{thr}=4 M_{\odot} pc^{-2}$ (thick dot-dashed lines), 
compared to observational data from various authors. }
\label{elfe_beta2_35_thr}
\end{figure*}

\begin{figure*}
\centering
\vspace{0.001cm}
\epsfig{file=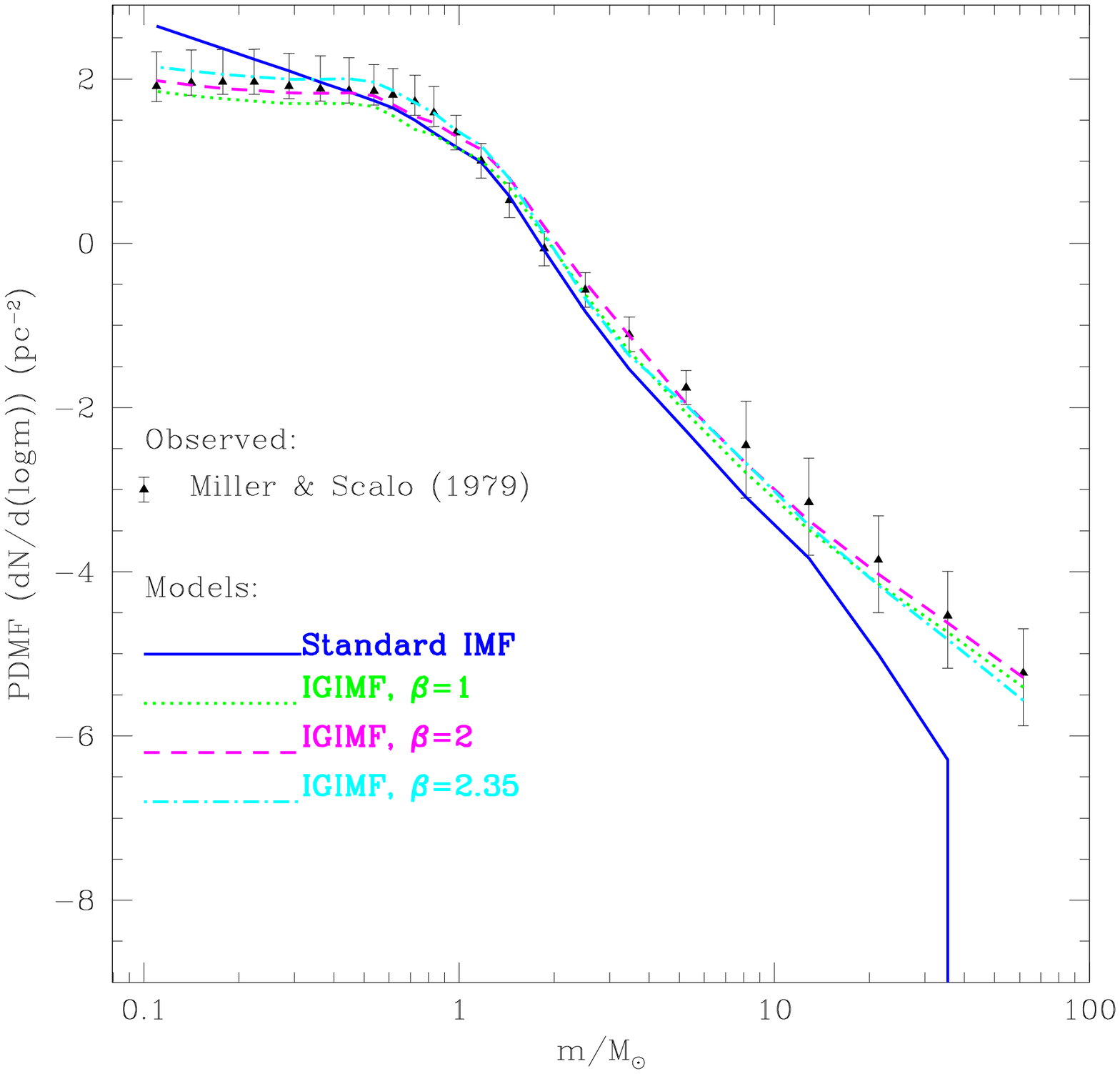,height=10cm,width=10cm}
\caption{Predicted present-day mass function computed by assuming the standard IMF (thick solid  line)
and the IGIMF with $\beta=1$ (thick dotted line), $\beta=2$ (thick dashed line) and $\beta=2.35$ (thick dot-dashed line). 
The observational PDMF is the one of  Miller \& Scalo (1979) (solid triangles with error bars). }
\label{pdmf}
\end{figure*}

\subsection{Effects of other parameters: the star formation threshold and $A_{Ia}$}
At this stage, it may be interesting to study the effects 
which other important parameters have on our results, such as the SF threshold and 
the type Ia SN realization probability $A_{Ia}$. 

As shown in Sect.~\ref{beta2}, the models which best reproduce the constraints considered in this paper 
are not sensitive to the adopted SF threshold. This is visible from the smooth behaviour of the 
calculated SF history, whereas we have seen that when the effect of the threshold is important, it 
presents an oscillatory behaviour, as happens for  the SFH of the model with the standard IMF. 
The case with $\beta=1$ is characterized by a high
 amount of mass restored by massive stars into the ISM, keeping its density always significantly above the threshold. 
Furthermore, in this case the SF efficiency which allows us to best reproduce the local features is lower than in the standard case. 
This causes the gas mass to be less efficiently consumed, so that the gas density level is always considerably above the threshold. 
The same is true for the case with $\beta=2$. 
In these two cases, halving  the  SF threshold to  $4 M_{\odot}/pc^2$ 
does not produce any effect on our results. On the other hand, the adoption of a significantly higher threshold, say $>10 M_{\odot}/pc^2$ 
would be at variance with the observations in local spiral discs (Martin \& Kennicutt 2001). 
The case with $\beta=2.35$ is the most sensitive to the assumption of the threshold. 
We show the effects of a threshold  $4 M_{\odot}/pc^2$ on the star formation history in Fig. ~\ref{snr_beta2_35_thr}. 
The decrease of the threshold causes the SF history to be smoother and to enhance
 gas consumption, therefore at the present time the galaxy model with 
$\Sigma_{thr}=4 M_{\odot}/pc^2$ presents a gas density and a stellar density 
slightly lower and higher than the model with $\Sigma_{thr}=7 M_{\odot}/pc^2$, respectively. 
On the other hand, the assumption of a lower threshold has no effect on the abundance ratios, and this can be seen in Fig.~\ref{elfe_beta2_35_thr}. 

It is worth noting that recent GALEX results 
indicate that the SF cutoff visible from the $H\alpha$ profiles of local star forming galaxies 
is instead absent in UV profiles (Boissier et al. 2007), suggesting that the existence of a SF threshold may 
be an observational selection effect, naturally explained by the IGIMF theory as shown by Pflamm-Altenburg \& Kroupa (2008). 
However, as shown by chemical evolution results, the SF threshold is fundamental in reproducing 
the metallicity gradients observed in the MW and in local galaxies (Chiappini et al. 2001; 2003), unless a variable star formation efficiency through the disc is assumed 
(Colavitti et al. 2009).  

It may be also interesting to discuss the possible effects of the type Ia SN realization probability $A_{Ia}$ on our results. 
A change in  $A_{Ia}$  has effects on the predicted type Ia SN rate, on the [X/Fe]-[Fe/H] plots and on the SMD plot, whereas 
it leaves unchanged any other quantity studied here. 
In the case $\beta=1$, the model which  best reproduces our set of 
observational constraints is  characterized by a SF efficiency $\nu=0.1$ Gyr$^{-1}$.  
To reproduce the abundance ratios as a function of [Fe/H], 
a higher value of $A_{Ia}$ would be required, so that all the curves would move downwards. 
This would improve also the fit to the observed stellar metallicity distribution, 
since an increase of the type Ia SN rate efficiciency would shift the computed SMD rightwards, probably in better agreement with the 
observations. However, as shown in Fig.~\ref{snr_beta1}, even with a  higher value for $A_{Ia}$, 
the observed local type II SN rate, the stellar mass  density and the gas density (which do not depend on $A_{Ia}$) would still not be reproduced. \\
We have seen that in the case $\beta=2$, no substantial modification of $A_{Ia}$ is required. \\
Finally,  in the case $\beta=2.35$, a lower value of 
 $A_{Ia}$ would be required to improve our match to the observations in the [X/Fe]-[Fe/H] plot, 
however this would shift the predicted SMD leftwards, exacerbating the discrepancy between the computed SMDs and the 
the observed distribution.

\subsection{The Present-day Mass Function}
\label{PDMF}
The present-day mass function (hereinafter PDMF) represents 
the mass function of living stars as observed in the solar neighbourhood. This quantity  
is an important diagnostic since 
it allows us to test,  beside the IMF, the star formation history of our best model, providing  further 
information on our parameters complementary to the ones previously discussed.

The number of stars formed in the time interval $(t,t+dt)$ and in the mass interval $(m,m+dm)$ is 
\begin{equation}
\xi(m,t) \psi (t) dm dt,
\end{equation}
(Tinsley 1980). The present number per unit mass  of stars with lifetime $\tau_{m} < T_{0}$ is 
\begin{equation}
n(m) = \int_{T_0-\tau(m)}^{T_0} \psi (t) \xi (m,t) \, dt.
\end{equation}
The total number of stars with mass between $M_{min}= 0.1 M_{\odot}$ and $m$ is  $N(m)$, 
calculated as 
\begin{equation}
N(m) = \int_{M_{min}}^{m} \int_{T_0-\tau(m')}^{T_0} \psi (t) \xi (m',t) \, dt \, dm',
\end{equation}
where $T_0$ is the present time, equal to 14 Gyr. 
The present-day mass function can be calculated as 
\begin{equation}
PDMF = dN(m)/d(log \,m).
\end{equation}
In Fig.~\ref{pdmf}, we show the local PDMF as estimated observationally by various authors, 
and as predicted by means of our models assuming the standard IMF 
and the IGIMF for all the three different cases for $\beta$, assuming the SF efficiencies of the best models 
as indicated by our fitness test described in Sect.~\ref{fitness}.  
The analytical PDMF of Chabrier (2003)  
has a  log-normal profile for masses $m \le 1 M_{\odot}$:
\begin{equation}
\frac{d \, N}{d \, (log \, m)}\propto \,exp\{-{(\log \, m\,\,-\,\,\log \, m_c)^2\over 2\,\sigma^2}\},
\end{equation}
with $m_c=0.079 M_{\odot}$ and $\sigma=0.69$, 
and for masses $m > 1 M_{\odot}$ it is given by a power-law:
\begin{displaymath}
     \frac{d \, N}{d \, (log \, m)}\propto \left\{ \begin{array}{l l}
                                      m^{-4.37} & 
			     \qquad {\mathrm{if}} \; 0 \le log(m/M_\odot) \le 0.54, \\
			              m^{-3.53} &
			     \qquad {\mathrm{if}} \; 0.54 \le log(m/M_\odot) \le 1.26, \\
                                     m^{-2.11} & 
			     \qquad {\mathrm{if}} \; 1.26 \le log(m/M_\odot) \le 1.80. \\
                                              \end{array} \right.
\end{displaymath}
The analytical PDMF of Kroupa et al. (1993) is a power-law:  
\begin{displaymath}
     \frac{d \, N}{d \, (log \, m)}\propto \left\{ \begin{array}{l l}
                                      m^{-1.3} & 
			     \qquad {\mathrm{if}} \; log(m/M_\odot) \le -0.3, \\
			              m^{-2.2} &
			     \qquad {\mathrm{if}} \; -0.3 \le log(m/M_\odot) \le 0, \\
                                     m^{-4.5} & 
			     \qquad {\mathrm{if}} \; 0 < log(m/M_\odot), \\
                                              \end{array} \right.
\end{displaymath}
The Kroupa et al. (1993) and Chabrier (2003) PDMF are very similar at masses $M<1 M_{\odot}$, 
and in the same mass range they are in very good agreement with the Miller \& Scalo (1979) PDMF. 
For masses $M>1 M_{\odot}$, no recent update exists and the reference measures are those of Miller \& Scalo (1979) and Scalo (1986), 
which are in very good agreement in this mass range. 
In the light of this, for purposes of clarity and to avoid confusion, as observational data in Fig.~\ref{pdmf} we plot 
only the PDMF as determined by Miller \& Scalo (1979). This is also the one used for the fitness test described in Sect.~\ref{fitness}.


From the analysis of Fig. ~\ref{pdmf}, interesting information can be drawn on the shape of the IMF and on the effects of 
other parameters such as the SF threshold. 
The PDMF computed with the standard IMF is in excellent agreement with the observational data 
in the range 0.4 $ M_{\odot}$  - 2  $M_{\odot}$. At stellar masses lower than 0.4 $ M_{\odot}$, the standard IMF is too steep and overestimates 
the observed number of low mass stars. On the other hand, the model computed with the standard IMF underestimates the observed 
distribution of massive stars with masses $>30  M_{\odot}$. 
This is due to the adoption of the SF threshold and its 
strong effects on the SF history of the solar neighbourhood at late times, inhibiting recent SF and hence causing the underabundance or absence 
of very massive stars.

The best models calculated with the IGIMF for $\beta=2$ and $\beta=2.35$ have different values for the SF efficiencies, but provide all similarly  
a very good fit to the observed PDMF.
 The model with $\beta=1$ slightly underestimates the low-mass end of the PDMF. \\
The small offsets among the computed distributions are due to different values adopted for the star formation efficiencies. 
These assumptions produce different values for the present-time stellar mass densities, which 
reflect into the small offsets visible in Fig. ~\ref{pdmf}. \\
Our results indicate that the values $\beta=2$  and $\beta=2.35$ should be preferred. 
In the case $\beta=2.35$, apparently the SF threshold does not affect the computed PDMF. 
The best model with the IGIMF and $\beta=2.35$ presents  no truncation in the PDMF since,  
even if the SFR has a gasping behaviour, the frequency of recent star formation events is sufficient 
to guarantee a significant presence of high-mass stars still living at the present time. \\
Finally, it is important to stress the importance of the fitness test used to determine which is 
the best model to reproduce all the obsevables, and how its results depend on the list of observables taken into account. 
In Table ~\ref{tab_fit}, we present our results for the fitness test performed considering (i) the whole set of observables and, for comparison, 
(ii) computed by considering all observables but the PDMF. 
In column 1, for each model we report the $\beta$ values. In column 2, we show the SF efficiencies. 
In column 3, we show the fitness obtained by using all the obsevables, and in column 4 we present the fitness computed by 
excluding the PDMF. It is clear that, once the PDMF is excluded and, 
if one considers the ability to reproduce all the chemical evolution constraints and 
physical properties such as the SN rates or the gas mass density, 
the two models with $\beta=2$, $\nu=0.3$ Gyr$^{-1}$ and $\nu=0.5$ Gyr$^{-1}$ are the best, but the standard model 
provides comparable results. 
However, the uncapability of the standard  model of accurately reproducing the observed PDMF has a strong impact on the fitness 
computed considering the whole set of observables. In this case, 
the two models with $\beta=2$ yield better results than the standard model. 

\subsection{The time variation of the IGIMF}

\begin{figure*}
\centering
\vspace{0.001cm}
\epsfig{file=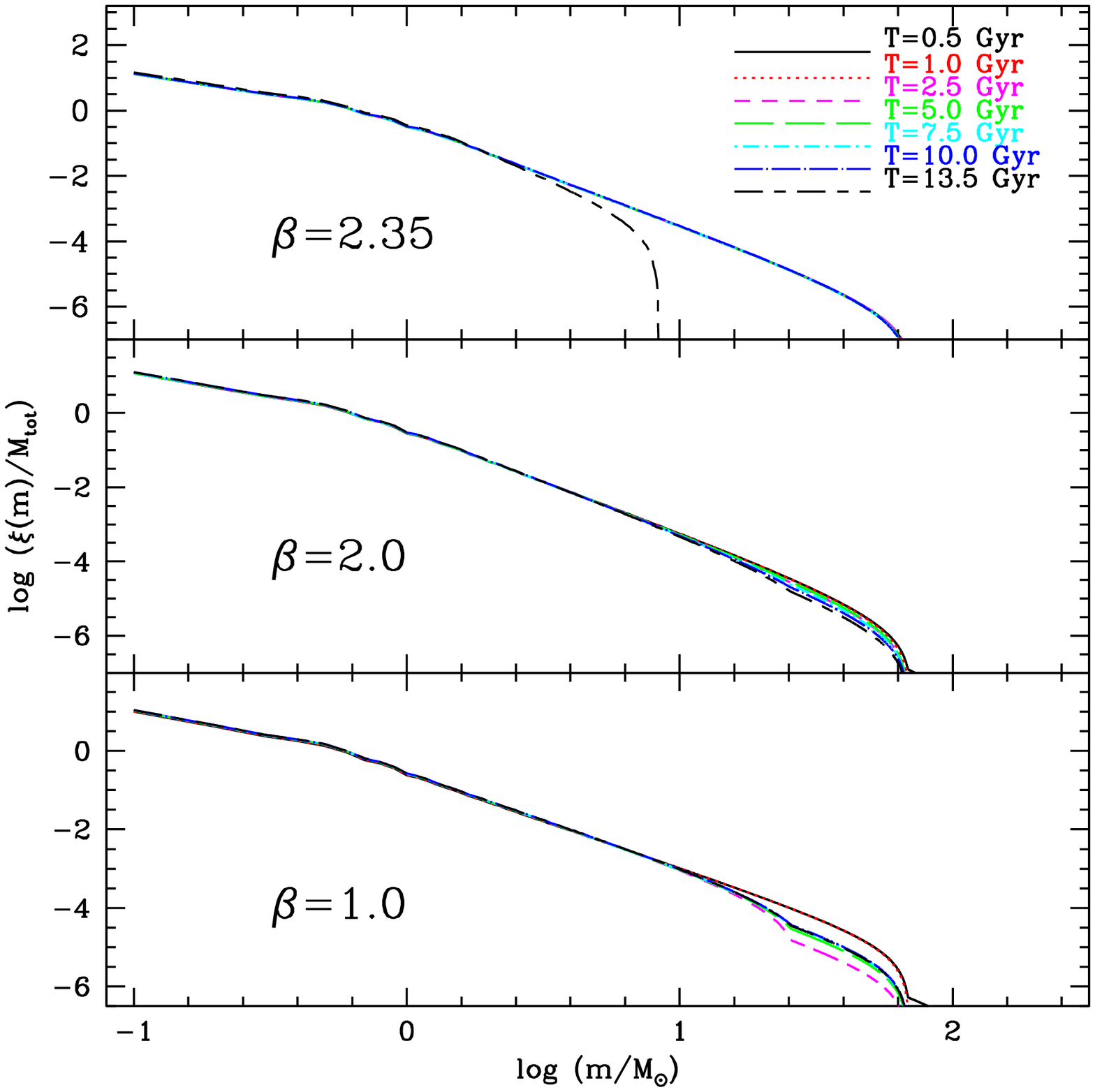,height=10cm,width=10cm}
\caption{Evolution of the IGIMF as a function of the cosmic time for the models  with $\beta=2.35$ and $\nu=1.5$ Gyr$^{-1}$ (upper panel), 
with $\beta=2$ and $\nu=0.5$ Gyr$^{-1}$ (central panel) and for the model with $\beta=1$ and $\nu=0.25$ Gyr$^{-1}$ (lower panel). 
In each panel, the curves represent IGIMFs at various times, calculated on the basis of the time variation of the star formation 
history of each model. 
}
\label{igimf_t}
\end{figure*}

In Fig.~\ref{igimf_t}, we show how the IGIMF varies as  a function of time 
in the case of the three best models, with $\beta=2.35$ and $\nu=2$ Gyr$^{-1}$ (upper panel), 
with $\beta=2$ and $\nu=0.5$ Gyr$^{-1}$ (central panel) and  with  $\beta=1$ and $\nu=0.1$ Gyr$^{-1}$ (lower panel).
In each panel, 
we show the IGIMF at various epochs, each one characterised by different SFR values.

The best model with $\beta=1$ is the one characterized by the lowest SF efficiency and presents a rather 
smooth SFH, although it has stronger variations than the other cases. 
In fact, as can be seen from Fig.~\ref{snr_beta1}, at time 1 Gyr the model with  $\nu=0.1$ Gyr$^{-1}$ has a SFR 
$<1 M_{\odot}/yr$, whereas at the present time it has a  SFR $\sim 2 M_{\odot}/yr$. 
On the other hand, in the period going from 1 Gyr to the present time, the other two models 
have SFR variations within a factor of 2, smaller than in the case with  $\nu=0.1$ Gyr$^{-1}$. 
The variations in the SFH of the best model are the cause  of 
the time variations in the IGIMF visible in Fig.~\ref{igimf_t} at the highest stellar masses (log$(M/ M_{\odot} \sim 1.4)$).

The best model with $\beta=2$ 
presents a rather smooth  SFR characterised by no strong variation in the period 1 Gyr-14 Gyr  (see Fig.\ref{snr_beta2}), 
not influenced by the effects of 
the star formation threshold.  This reflects the very small variation of the IGIMF with cosmic time. \\
The threshold plays an important role in the evolution of the model with $\beta=2.35$, 
whose SFH presents a gasping behaviour throughout the whole cosmic time. 
As a consequence, the IGIMF presents strong variations.

Our results show that within the IGIMF theory, if the absolute best model  reproducing the properties of the solar neighbourhood 
is the one with $\beta=2$ and $\nu=0.5$ Gyr$^{-1}$, 
the variation of the IMF with time must be  small. 

In principle, other objects with extreme variations in the SFH such as starburst galaxies should exhibit 
a strongly time-varying IMF. This will be an interesting subject for future work.

\begin{table*}
\vspace{0cm}
\begin{flushleft}
\caption[]{Fitness test for various models in two different cases. 
In columns 1 and 2, for each model the value of $\beta$ and SF efficiency values are reported, respectively. 
In columns 3 and 4, the results of the fitness test computed by considering the whole set of observables and 
all observables but the PDMF are shown, respectively. }
\begin{tabular}{ll|ll}
\noalign{\smallskip}
\hline
\hline
\noalign{\smallskip}
 $\beta$             &   $\nu$          & Fitness,              & Fitness             \\ 
                     &                  & all observables       & all obs. except PDMF   \\ 
\noalign{\smallskip}
\noalign{\smallskip}
\hline
\noalign{\smallskip}
 1.00 & 0.10  & 0.54  &  0.56  \\   
 1.00 & 0.25  & 0.53  &  0.57    \\   
 1.00 & 0.50  & 0.53  &  0.55     \\
 2.00 & 0.10  & 0.47  &  0.51    \\
 2.00 & 0.30  & 0.66  &  0.69    \\
 2.00 & 0.50  & 0.68  &  0.72    \\
 2.35 & 1.00  & 0.48  &  0.53    \\
 2.35 & 1.50  & 0.46  &  0.55    \\
 2.35 & 2.00  & 0.52  &  0.55    \\
\hline         
 Standard & 1.00  & 0.56  &  0.68    \\
\hline
\hline
\end{tabular}
\label{tab_fit}
\end{flushleft}
\end{table*}

\section{Conclusions}

The initial mass function regulates the number of stars of different intial mass born per stellar generation, hence 
it plays a fundamental role 
in galactic chemical evolution studies.  
The aim of this paper was to test the effects of adopting the integrated galactic initial mass function (IGIMF) on the chemical evolution of 
the solar neighbourhood. 
The IGIMF is computed from the combination 
of the stellar intial mass function, i.e. the mass function in single star clusters, and the embedded cluster mass function, i.e. 
a power law with index $\beta$, and taking into account that 
within each single cluster, the maximum stellar mass is a function of the total mass of the cluster. The result is a time-varying IMF which 
is a function of the galactic star formation rate.  
We applied the formalism developed by Weidner \& Kroupa (2005) to a chemical evolution model for the solar neighbourhood, based on 
the two-infall model by Chiappini et al. (1997). 
For the embedded cluster mass function, we tested three different values of $\beta$ and studied 
various physical quantities  and abundances for various chemical elements, comparing our results to the ones obtained by means of a standard 
IMF, constant in time and similar to the Scalo (1986) IMF. A statistical test to determine which is the best model in reproducing 
the set of observational data considered in this work has been developed. 
Our results can be summarized as follows:\\
1) The value of $\beta$ has important effects on the predicted star formation history. Also the effects of the SF threshold may be weaker 
or stronger, depending on the value chosen for  $\beta$. 
In general, lower absolute values of $\beta$ imply a flatter IGIMF, hence a larger number of massive stars and larger mass ejection rates. 
This translates into a larger amount of  gas available for star formation and gas density values never below  
the threshold and a smooth star fomation history. \\
2) The value of $\beta$ has an obvious deep impact on the predicted SN rates.   
Beside this, also other quantities can be influenced by this parameter. 
In general,  a lower  $\beta$ implies higher mass ejection rates from massive stars, 
hence at any time a  larger gas mass density. \\
3) The interstellar abundances are strongly influenced by the parameter $\beta$. 
In general, lower absolute values of $\beta$  imply higher metallicities and higher 
[O/Fe] values at a given metallicity. \\
4) We have considered several chemical evolution constraints, including the observed 
local SN rates, local gas and stellar surface densities, 
the abundance ratios in local stars and the stellar metallicity distribution and, by varying the SF efficiency, we   
tested which assumption 
for the embedded cluster mass function exponent $\beta$ provides the best simultaneous match to all of these observables. 
Our fitness test indicates that  
the best result has been obtained by assuming $\beta =2 $, 
which produces an IMF resembling that of our standard model and that 
allows us to account for most of the  observables.\\
5) Useful hints on the initial mass function come from the study of the present day mass function, 
i.e. the mass function of living stars observed today in the solar neighbourhood. 
The PDMF is a very good test of the chosen star formation history (Elmegreen \& Scalo 2006) once an IMF has been assumed and 
a valuable test for the IMF once the SFR is assumed. 
In the case of the standard IMF, the star fomation history assumed here predicts a present-time SFR in good agreement with the observations, 
and is based on the results by Kennicutt (1989). 
Models with the IGIMF and different assumptions for $\beta$ provide different results. 
Lower values for $\beta$ produce a PDMF richer in massive stars, whereas 
higher values of  $\beta$ imply  
steeper IGIMFs and a lower relative number of massive stars. The model with $\beta =2 $ should be considered 
the best since it allows to reproduce at the same time the observed PDMF and to account for most of the chemical evolution 
constraints considered in this work.

In the future, we plan to extend our study of the effects of the IGIMF in galaxies of different morphological types. 
As forthcoming step, we will investigate how  
the IGIMF affects the chemical evolution of  dwarf galaxies.

\section*{Acknowledgments}
We thank C. Chiappini for several interesting discussions.

\label{lastpage}


\begin{thebibliography}{99}
\bibitem[]{} Bensby, T.; Feltzing, S.; Lundstr\"om, I., 2004, A\&A, 415, 155
\bibitem[]{} Boily, C. M.; Kroupa, P., 2003a, MNRAS, 338, 665 
\bibitem[]{} Boily, C. M.; Kroupa, P., 2003b, MNRAS, 338, 673
\bibitem[]{} Boissier, S., et al., 2007,  ApJS, 173, 524
\bibitem[]{} Calura, F., Matteucci, F., 2006a, ApJ, 652, 889
\bibitem[]{} Calura, F.; Pipino, A.; Chiappini, C.; Matteucci, F.; Maiolino, R., 2009, A\&A, 504, 373
\bibitem[]{} Cappellaro, E., 1996, in, eds, Proc. IAU Symp. 171, New Light on Galaxy Evolution. Kluwer Academic, Dordrecht, p.81
\bibitem[]{} Cayrel, R., Depagne, E., Spite, M., et al. 2004, A\&A, 416, 1117
\bibitem[]{} Cescutti, G., 2008, A\&A, 481, 691
\bibitem[]{} Chabrier, G., 2003, PASP, 115, 763 
\bibitem[]{} Chiappini, C., Matteucci, F., Gratton, R. 1997, ApJ, 477, 765
\bibitem[]{} Chiappini, C., Matteucci, F., Padoan, P. 2000, ApJ, 528, 711
\bibitem[]{} Chiappini,  C., Matteucci,  F., Romano,  D., 2001, ApJ, 554, 1044
\bibitem[]{} Chiappini, C.; Romano, D.; Matteucci, F., 2003, MNRAS, 339, 63
\bibitem[]{} Colavitti, E.; Cescutti, G.; Matteucci, F.; Murante, G. 2009, A\&A, 496, 429
\bibitem[]{} Dame,  T. M., 1993, in  Holt  S. S., Verter  F., eds, AIP Conf. Proc. 278, ``Back to the Galaxy'', Am. Inst. Phys.,  New York  , p. 267
\bibitem[]{} Elmegreen, B. G., Scalo, J., 2006, ApJ, 636, 149
\bibitem[]{} Fran\c cois, F.,  Matteucci, F., Cayrel, R., et al. 2004, A\&A, 421, 613 
\bibitem[]{} Gilmore, G., Wyse, R., Kuijen, K., 1989, in ``Evolutionary Phenomena in Galaxies'', ed. J. Beckman, B. Pagel (Cambridge: Cambridge University Press), 172
\bibitem[]{} Gratton, R., Carretta, E., Matteucci, F., Sneden, C., 2000, A\&A, 358, 671
\bibitem[]{} Grevesse, N., Asplund, M., Sauval, A. J. 2007, Space Sci. Rev., 130, 105
\bibitem[]{} Holmberg, J.; Flynn, C., 2004, MNRAS, 352, 440
\bibitem[]{} Israelian, G., Ecuvillo, A., Rebolo, R., et al. 2004, A\&A, 421, 649
\bibitem[]{} Iwamoto, K., Brachwitz, F., Nomoto, K., et al. 1999, ApJS, 125, 439
\bibitem[]{} Jorgensen, B. R. 2000, A\&A, 363, 947 
\bibitem[]{} Kennicutt, R. C., 1989, ApJ, 344, 685 
\bibitem[]{} Kennicutt, R. C., 1998, ApJ, 498, 541
\bibitem[]{} Kroupa, P., Weidner, C. 2003, ApJ, 598, 1076
\bibitem[]{} Kroupa, P., Tout, C. A., Gilmore, G. 1993, MNRAS, 262, 545
\bibitem[]{} Kulkarni, S. R., Heiles, C. 1987, in Interstellar Processes, ed. D. Hollenbach,  H. Thronson (Dordrecht: Kluwer), 87
\bibitem[]{} Lada, C.J., Lada, E.A., 1991, in Janes K. ed., ASP Conf. Ser. Vol. 13, The Formation and Evolution of Star Clusters, Astron. Soc. Pac. San Francisco, p. 3
\bibitem[]{} Lada, C. J.,  Lada, E. A. 2003, ARA\&A, 41, 57
\bibitem[]{} Larsen, S. S., Richtler, T. 2000, A\&A, 354, 836
\bibitem[]{} Larson, R. B. 1976, MNRAS, 176, 31
\bibitem[]{} Martin, C. L.; Kennicutt, R. C., 2001, ApJ, 555, 301
\bibitem[]{} Massey, P., Hunter, D. A. 1998, ApJ, 493, 180 
\bibitem[]{} Matteucci, F., Fran\c cois P., 1989, MNRAS, 239, 885
\bibitem[]{} Matteucci,  F., Greggio,  L., 1986, A\&A, 154, 279
\bibitem[]{} Matteucci, F., Recchi, S. 2001, ApJ, 558, 351
\bibitem[]{} Matteucci,  F., 2001, ASSL Vol. 253, The Chemical Evolution of the Galaxy. Kluwer,  Dordrecht,  293
\bibitem[]{} Matteucci, F., Panagia, N., Pipino, A., Mannucci, F., Recchi, S., Della Valle, M., 2006, MNRAS, 372, 265
\bibitem[]{} Matteucci, F.; Spitoni, E.; Recchi, S.; Valiante, R., 2009, A\&A, 501, 531
\bibitem[]{} M\'era, D., Chabrier, G., Schaeffer, R. 1998, A\&A, 330, 937
\bibitem[]{} Miller, G. E.; Scalo, J. M., 1979, ApJS, 41, 513
\bibitem[]{} Nordstr\"om, B., Mayor, M.; Andersen, J.; Holmberg, J.; Pont, F.; J\o rgensen, B. R.; Olsen, E. H.; Udry, S.; Mowlavi, N., 2004, A\&A, 418, 989
\bibitem[]{} Olling, R. P.; Merrifield, M. R., 2001, MNRAS, 326, 164
\bibitem[]{} Palou{\v s}, J., \& Theis, C.\ 2007, A\&A, 461, 155 
\bibitem[]{} Pflamm-Altenburg, J.; Kroupa, P., 2008, Nature, 455, 641
\bibitem[]{} Pflamm-Altenburg, J., Weidner, C., Kroupa, P. 2007, ApJ, 671, 1550
\bibitem[]{} Rana, N. 1991, ARA\&A, 29, 129
\bibitem[]{} Recchi, S.; Calura, F.; Kroupa, P., 2009, A\&A, 499, 711
\bibitem[]{} Romano, D., Matteucci, F., Salucci, P., Chiappini, C. 2000, ApJ, 539, 235
\bibitem[]{} Romano, D., Chiappini, C., Matteucci, F., Tosi, M., 2005, A\&A, 430, 491
\bibitem[]{} R{\u u}{\v z}i{\v c}ka, A., Palou{\v s}, J., Theis, C.\ 2007, A\&A, 461, 155 
\bibitem[]{} Salpeter, E. E. 1955, ApJ, 121, 161
\bibitem[]{} Samland, M., Hensler, G., Theis, C. 1997, ApJ, 476, 544
\bibitem[]{} Scalo, J. M., 1986, FCPh, 11, 1
\bibitem[]{} Sommer-Larsen, J.; G\"otz, M.; Portinari, L., 2003, ApJ, 596, 47
\bibitem[]{} Spite, M., Cayrel, R., Plez, B., et al. 2005, A\&A, 430, 655
\bibitem[]{} Talbot, R. J.; Arnett, W. D., 1971, ApJ, 170, 409
\bibitem[]{} Theis, Ch.; Kohle, S., 2001, A\&A, 370, 365  
\bibitem[]{} Tinsley, B. M. 1980,  FCPh, 5, 287
\bibitem[]{} Van den Hoeck, L. B. Groenwegen, M. A. T., 1997, A\&AS, 123, 305
\bibitem[]{} Weber, M.; de Boer, W., 2010, A\&A, 509, 25
\bibitem[]{} Weidner, C., Kroupa, P. 2005, ApJ, 625, 754
\bibitem[]{} Weidner, C., Kroupa, P. 2006, MNRAS, 365, 1333
\bibitem[]{} Weidner, C., Kroupa, P., Larsen, S.S. 2004, MNRAS, 350, 1503
\bibitem[]{} Whelan, J., Iben, I. Jr. 1973, ApJ, 186, 1007 
\bibitem[]{} Zhang, Q., Fall, S. M. 1999, ApJ, 527, L81 
\end{thebibliography}
\end{document}